# A New Heterodyne Megasonic Piezoresponse Force Microscopy with High-frequency Excitation beyond 100 MHz

Qibin Zeng[1], Hongli Wang[1,3], Qicheng Huang[2], Zhen Fan[2], Kaiyang Zeng[1,*]

[1] Department of Mechanical Engineering, National University of Singapore, Singapore 117576, Singapore

[2] Institute for Advanced Materials, South China Academy of Advanced Optoelectronics, South China Normal University, Guangzhou 510006, China

[3] Guangdong Institute of New Materials, National Engineering Laboratory for Modern Materials Surface Engineering Technology, The Key Lab of Guangdong for Modern Surface Engineering, Guangzhou, 510651, China

*Correspondence and requests for materials should be addressed to K. Z. (email: mpezk@nus.edu.sg)

ABSTRACT

Piezoresponse Force Microscopy (PFM), as a powerful nanoscale characterization technique, has been extensively utilized to elucidate diverse underlying physics of ferroelectricity. However, the intensive study of conventional PFM has revealed a growing number of concerns and limitations which are largely challenging its validity and application. Herein, we developed a new advanced PFM technique, named Heterodyne Megasonic Piezoresponse Force Microscopy (HM-PFM), which uniquely uses $10^6$ to $10^8$ Hz high-frequency excitation and heterodyne method to measure the piezoelectric strain at nanoscale. We report that HM-PFM can unambiguously provide standard ferroelectric domain and hysteresis loop measurements, and an effective domain characterization with excitation frequency up to ~110 MHz has been realized. Most importantly, owing to the high-frequency and heterodyne scheme, the contributions from both electrostatic force and electrochemical strain can be significantly minimized in HM-PFM. Furthermore, a special difference-frequency piezoresponse frequency spectrum (DFPFS) measurement is developed on HM-PFM and a distinct DFPFS characteristic is observed on the materials with piezoelectricity. It is believed that HM-PFM can be an excellent candidate for the piezoelectric or ferroelectric studies where the conventional PFM results are highly controversial.

With the growing demands of high density, miniaturization and high integration for devices, micro- and nanoscale ferroelectric structures or phenomena have attracted extensive attention from fundamental science to applications.[1, 2] Although numerous theories have already been established for ferroelectrics at macroscale, many essential mysteries with respect to micro- or nanoscale ferroelectric behaviors, such as the polarization dynamics, domain growth kinetics and the surface-screening mechanisms, still remain ambiguous and pending to be elucidated.[3, 4] Piezoresponse Force Microscopy (PFM), as an important branch of Scanning Probe Microscopy (SPM), has evolved as the mainstream technique towards unveiling these underlying ferroelectric physics since its inception in 1992.[2-7] In recent years, however, the extensive application of PFM has revealed a growing number of challenges and concerns about this technique, which is greatly challenging its validity in many ferroelectric studies recently.[6, 8, 9] Among these concerns, the signal source issue is pressingly pending to be addressed because the source of the signal is of fundamental importance for reaching correct interpretation of the PFM results.[6, 8, 9]

Electrostatic force is one of the most intractable issues which continuously influences the PFM results since its invention. Because the existence of the electrostatic force can give rise to significant artifacts or misinterpretations in PFM-based ferroelectric studies, resulting large numbers of ferroelectric-like observations in numbers of non-ferroelectric materials.[6, 8-15] Therefore, a large amount of research work has been implemented to eliminate or quantify the electrostatic force contribution in PFM measurements.[2, 9-13, 16-25] It has been proved that the electrostatic contribution can be relatively minimized by various methods, such as imaging the materials with strong piezoelectricity, using stiff or shielded probes and applying DC compensation voltage.[9, 11, 17, 19-21, 26] Furthermore, considering the difficulties in technically eliminating the electrostatic force, off-line analysis strategies or complementary experiments (such as contact Kelvin Probe Force Microscopy) are proposed to quantify the electrostatic force contribution thus achieving a correct interpretation of the PFM results.[10, 23, 24, 27] However, each of these methods is subjected to specific limitations,[12, 21, 22, 26, 28, 29] and so far more effective approaches respect to addressing the electrostatic force issue are still highly necessitated. In addition, in some materials, the electrochemical Vegard strain caused by the diffusion and electromigration of the mobile ions can also affect the PFM measurements significantly.[8, 9, 12] Due to the similarity of the electrochemical and piezoelectric strains, it is difficult to differentiate the electrochemical and piezoelectric signals especially if the property of the sample is unknown beforehand, i.e., whether the material is piezoelectric or not.[6, 13, 30-34] Although attentions have been paid by researchers to identify the true PFM signal,[8, 33, 35] these approaches are restricted by the materials and the co-existence of electrostatic force during the measurements.[10-12] By far, attributing the signal source of PFM to piezoelectric or electrochemical strains mainly relies on the material's property, and the effective ways to reduce the influence of electrochemical strain are still pending to be explored.[13] Except for the electrostatic force and electrochemical strain, the electrostrictive coupling also contributes to the PFM signal.[33, 36, 37] Since the electrostriction shows a quadratic relationship with respect to the applied electric field,[38, 39] its influence on PFM is similar to that of the electrostatic force. Moreover, there may be multiple other electric field-induced effects which can directly or indirectly induce responses in PFM measurements, including electrochemical dipoles, charge injection, field effect, electrochemical reactions, flexoelectricity and Joule heating and so on.[6, 8, 9, 15, 16, 33, 37, 39-43] Obviously, the issue on signal sources causes large complexity and uncertainty in current PFM measurements, which greatly challenges its application especially for the materials with unknown properties. Even though huge effort has been made to purify the PFM signal, the inherent complexity of electrochemical coupling phenomena still pressingly requires further substantial progresses to achieve an ideal PFM measurement.

Recently, high-frequency PFM working at high eigenmodes of the cantilever has been put forward to effectively minimize the electrostatic force contribution.[18, 20, 25] But the associated decrease of detection sensitivity, laser spot size effect, large bandwidth requirement for photodetector and lock-in amplifier make the application of high-frequency PFM very limited.[18, 25] Despite high-frequency PFM has seldom received attentions due to the current technical restrictions, using high-frequency excitation to detect piezoelectric strain does provide a meaningful instruction for the development of advanced PFM methods. In this study, by performing an overall assessment of using high-frequency excitation in PFM, it is found that several substantial improvements can be realized simultaneously, including minimizing the electrostatic force-induced cantilever vibration, attenuating electrochemical Vegard strain and electrostriction effects as well as reducing the influence of dynamic electrochemical processes. Most importantly, to effectively break the technical limitations and realizing PFM measurements at much higher frequency, heterodyne detection scheme has been introduced to PFM. Based on using high-frequency excitation and heterodyne detection, a new advanced PFM technique which focuses on detecting high-frequency piezoelectric strain is developed here. This technique, that we name as ***Heterodyne Megasonic Piezoresponse Force Microscopy*** (***HM-PFM***),[44] utilizes both electrical and mechanical drives with MHz frequency to probe the surface piezoelectric vibration at nanoscale via heterodyne detection method. Conventional ferroelectric materials are used to test the basic ferroelectric domain characterization, switching spectroscopy and high-frequency operation capabilities of the HM-PFM. The influence of electrostatic force in the HM-PFM measurement is systematically studied and compared with the results from conventional PFM. The results unambiguously reveal that HM-PFM can obtain ideal ferroelectric domain image, standard ferroelectric hysteresis loop and operate up to as high as ~110 MHz, at the same time, the electrostatic force contribution has been significantly minimized. Finally, the special measurement offered by HM-PFM, the difference-frequency piezoresponse frequency spectrum (DFPFS),[44] is demonstrated on three types of functional materials including dielectric, Lithium-ion battery and ferroelectric materials. The DFPFS results do show a remarkable difference between the piezoelectric and non-piezometric materials tested here. At the same time, the results also indicate that the electrochemical strain has been considerably attenuated in HM-PFM.

## Results

**High-frequency excitation.** In HM-PFM, the excitation frequency for tip and sample are both located within high-frequency region, typically $10^6$ Hz to $10^8$ Hz. As mentioned above, to operate PFM at high frequency does provide substantial improvement for numbers of signal source issues, which include minimizing the electrostatic force-induced cantilever vibration, attenuating electrochemical Vegard strain and electrostriction effects as well as reducing the influence of dynamic electrochemical processes, and bellow are detail analyses.

The electrostatic force-induced cantilever vibration can be significantly minimized when sample is excited by high-frequency electric filed. For the frequently used free rectangular AFM cantilever, the flexural vibration in air can be described by the classical Euler-Bernoulli beam theory, the effective force constant $k_n$ and quality factor $Q_n$ of $n^{th}$ eigenmode are respectively given by (ignoring the internal damping):[45]

$$k_n = \frac{\xi_n^4}{4} \frac{EI}{L^3} \tag{1}$$

$$Q_n = \frac{\sqrt{mEI}}{c} \left(\frac{\xi_n}{L}\right)^2 \tag{2}$$

where $E$, $I$, $L$ and $m$ are Young's modulus, moment of inertia, length and mass per unit length of the cantilever, respectively; $c$ is the hydrodynamic damping coefficient and $\xi_n$ is the wave numbers ($\xi_1$ =1.875, $\xi_2$ =4.694…) of the infinite flexural vibration modes which increases linearly with the mode number.[46] Equations (1) and (2) indicate that the effective force constant $k$ will increase with the 4$^{th}$ power of the mode number, while quality factor $Q$ only increase quadratically with the mode number (see Supplementary Fig. 1). Therefore, even ignoring the internal damping, the excitation of high eigenmodes will become more and more difficult. A detail calculation for the amplitude response of AFM cantilever as a function of excitation frequency has been performed and presented in the Supplementary Information (Supplementary Note 1). The results clearly show that the oscillation amplitude decays dramatically with increasing frequency, implying that the electrostatic force induced cantilever vibration will be significantly minimized if PFM is operated at high frequency.

With high-frequency excitation, the contribution from the Electrochemical Vegard strain to PFM results can also be largely reduced. The Vegard strain is resulted from the tip voltage-induced diffusion and electromigration of mobile ions. Since the motion of ion is highly frequency dependent, it is possible to change the magnitude of Vegard strain by changing the excitation frequency. In a typical PFM operation frequency range (~300 kHz), the relationship between Vegard strain caused surface displacement and tip voltage frequency is employed as follows[35, 47]

$$u_3 = \frac{2(1+v)\beta\sqrt{D}}{\eta} \frac{V_{ac}}{\sqrt{\omega}} \tag{3}$$

where $u_3$, $v$, $\beta$, $D$, $V_{ac}$ and $\omega$ are the amplitude of the surface displacement, Poisson's ratio, effective Vegard coefficient, the diffusion coefficient of the ion, amplitude and frequency of AC voltage, respectively, and $\eta$ represents the linear relation between chemical potential and applied electric field. According to Equations (3), the surface displacement is proportional to the reciprocal of the square root of the frequency, implying that the Vegard strain effect can be effectively suppressed at high frequency.[13, 48] Furthermore, if the modulation frequency is very high, i.e., far higher than the cut-off frequency of the ionic diffusion, the electrochemical Vegard strain is expected to become negligible, as the ions cannot diffuse as fast as the applied AC voltage (i.e., the ions are in a quasi-static state).[13, 48, 49] Both theoretical and experimental results have already demonstrated the attenuation of Vegard strain with increasing frequency for several lithium-containing materials.[50, 51]

In addition, increasing the frequency of probing wave can decrease the influences from electrostriction and dynamic electrochemical processes in the PFM measurement as well. Considering the deformation along the direction of applied electric field $E_{elec} = E_0 + E_{ac}\sin(\omega t)$, the electrostriction strain $S = Q_{es}(P_0 + \varepsilon E_{elec})^2$ and the first-harmonic strain which contributes to the piezoresponse signal is $S_\omega = 2Q_{es}(\varepsilon P_0 E_{ac} + \varepsilon^2 E_0 E_{ac})\sin(\omega t)$, where $Q_{es}$ is the polarization related electrostrictive coefficient, $P_0$ is an static net polarization (e.g. spontaneous polarization) and $\varepsilon$ is the dielectric constant.[37, 38] It has been found that the electrostrictive effect is usually governed by the dielectric constant instead of electrostrictive coefficient, and large electrostriction mainly exists in the materials with pretty large dielectric constant.[52, 53] According to the Debye relaxation-based dielectric dispersion relationship, the dielectric constant (real part) will gradually decrease with increasing frequency.[54] Therefore, increasing the frequency of the electric field, the electrostrictive strain will attenuate due to the decrease of dielectric constant. Furthermore, some other dynamic electrochemical processes, such as the formation of defects under the sample surface, water-splitting reaction, nano-oxidation, surface redox reaction and certain unusual electrochemical phenomena may also affect the PFM measurement especially in the ambient environment.[8, 41, 42, 55, 56] Typically, the majority of these electrochemical processes are dominated by the motion of the ions, defects and vacancies, etc. Since the motion of these particles have a specific temporal scale, it is safe to predict that once the motion of the particles cannot follow the variation of the drive voltage, the associated dynamic electrochemical processes will be largely suppressed.[41, 50]

**Heterodyne detection.** As the direct high-frequency detection is subjected by multiple limitations as precious discussion, we therefore introduce an indirect scheme, the heterodyne detection, to break those limitations. Heterodyne detection is an extensively used method in the measurement and analysis of high-frequency signals. The heterodyne process is to down-convert the high-frequency signal to a lower, easily measurable frequency by mixing it with a known reference signal.[57] In multi-frequency SPM, the heterodyne detection plays an essential role in the measurement of high-frequency signals since the direct high-frequency detection is usually difficult or impractical in SPM.[58, 59] To realize heterodyne detection in PFM, a high-frequency reference is required. In the designed HM-PFM here, the reference is provided by mechanically driving the cantilever via the holder transducer. Considering both tip and sample surface have a vibration, $A_t\sin(2\pi f_t t + \phi_t)$ and $A_s\sin(2\pi f_s t + \phi_s)$ respectively, the time-dependent tip-sample interaction force now is given by:

$$F_{ts}(z,t) = F_{ts}\left[z_0 + A_t\sin(2\pi f_t t + \phi_t) - A_s\sin(2\pi f_s t + \phi_s)\right] \tag{4}$$

where $z$ and $z_0$ are the instantaneous and equilibrium tip-sample separations, respectively. Since the tip-sample interaction force $F_{ts}$ depends non-linearly with $z$, under the condition of small vibration amplitudes, $F_{ts}$ can be approximately expressed by a Taylor series at $z = z_0$ up to second order:

$$F_{ts}(z,t) \approx F_{ts}(z_0) + F'_{ts}(z_0)(z-z_0) + \frac{1}{2}F''_{ts}(z_0)(z-z_0)^2 \tag{5}$$

Combining Equation (4) and (5) and ignoring all the static and high-frequency items, the difference-frequency force is given by (see details in Supplementary Note 2):

$$F_{ts}(z)_{diff} = -\frac{1}{2}F_{ts}''(z_0)A_t A_s \cos\left[2\pi f_{diff}t + \phi_s - \phi_t\right] \tag{6}$$

in which $f_{diff} = f_s - f_t$. If the drive frequencies of tip and sample are set to be close, the difference-frequency force $F_{ts}(z)_{diff}$ will be located at low-frequency range and can normally drive the cantilever to oscillate and then detected by conventional SPM set-up, even though very high frequencies are used. For the purpose of ferroelectric characterization, revealing the contrasts of sample vibration amplitude $A_s$ and phase $\phi_s$ are necessary because the information of interest (such as domain wall and polarization direction) is included in $A_s$ and $\phi_s$. From Equation (6), it is clear to see if the 2$^{nd}$-order force gradient, tip vibration amplitude and phase are constant, the sample vibration amplitude and phase can be extracted via detecting the difference-frequency cantilever vibration (see more details in Supplementary Note 2). As the tip always keeps a constant force on sample surface during contact scanning, the above-mentioned requirements can be achieved especially when scanning area is flat and uniform. Therefore, by using heterodyne detection, the limitations in previous high-frequency PFM have been removed and the characterization of ferroelectric domain with very high excitation frequency now become possible (as the ~110 MHz realized in this study).

**Experimental set-up.** In this study, the model HM-PFM system is constructed on a commercial SPM system (SPA400, Seiko Instruments Inc.). The complete set-up of this model system is depicted in. Fig. 1. In the HM-PFM designed here, tip is set to be grounded while the drive signal for stimulating electromechanical strain is sent to the conductive substrate of the sample. In order to mechanically drive the cantilever at high frequency, the original probe holder is specially modified to enable an effective external high-frequency excitation. During measurement, the tip vibration is stimulated by the holder transducer via holder drive $V_{holder}$ at frequency $f_t$, while and the sample vibration is excited by the sample drive $V_s$ at frequency $f_s$. The difference-frequency oscillation (here we call it difference-frequency piezoresponse, DFP) generated from the heterodyne process is detected by the position sensitive detector (PSD) and then demodulated by the lock-in amplifier. A home-made analog multiplier and a low-pass filter (LPF) are used to produce difference-frequency reference signal for coherent demodulation. Alternatively, the reference signal can be provided internally by synchronizing the clocks of signal source and lock-in amplifier.[44] The demodulation results, i.e. amplitude and phase, of the DFP signal are sent to the controller of AFM and then synchronously imaged with the topography (all the amplitude images here are shown in dimension of a.u.). In order to enhance the signal-to-noise ratio (SNR) of the DFP signal, the tip drive frequency $f_t$ is typically set to close the eigenfrequency of a high eigenmode (typically in MHz range) to generate a stronger vibration; the difference frequency $f_{diff}$ is usually set to a value near the 1st or 2nd-order contact resonance frequency of the cantilever, where the resonance amplification can be utilized;[58] then the sample drive frequency $f_s$ is set to $f_t + f_{diff}$. The determination of $f_t$ and $f_s$ are completed via a self-developed frequency-selecting program which typically works within 1 MHz to 200 MHz.

**Ferroelectric domain characterization and switching spectroscopy.** In this study, the periodically poled lithium niobate (PPLN) single crystal is used as the standard test sample, similar to other PFM characterization.[28, 60] Fig. 2a-c shows the typical scanning results of the PPLN by HM-PFM with the operating frequencies $f_t$ = 8.356 MHz, $f_s$ = 8.63959 MHz ($f_{diff}$ = 283.59 kHz) and other scanning parameters are nearly the same with that used in conventional PFM.[61] From the amplitude and phase images of the DFP signal, the domain walls between the two adjacent domains can be clearly observed in the amplitude image (Fig. 2b), and the periodical domains with alternative upward and downward polarization are distinctly revealed in the phase image (Fig. 2c). Meanwhile, Fig. 2b and 2c indicate a uniform amplitude distribution and a nearly 180° phase difference between the domains with opposite polarization, which agrees well with the characteristics of the proposed "ideal" PFM measurements of ferroelectric materials.[28] An additional ferroelectric material, Pb(Zn$_{1/3}$Nb$_{2/3}$)O$_3$-9%PbTiO$_3$ (PZN-9%PT) single crystal, is also studied here to test the HM-PFM. The PZN-9%PT sample tested here has spontaneously polarized domains in which the polarization is along thickness direction, and the upward and downward domains are randomly distributed.[62] Typical topography and simultaneously obtained HM-PFM results of the PZN-9%PT sample are shown in Fig. 2d-f, where the operation frequencies are: $f_t$ = 8.395 MHz, $f_s$ = 8.6813 MHz ($f_{diff}$ = 286.3 kHz). Clear labyrinthine domain pattern and ~180° phase difference between the upward and downward domains can be observed, which are highly consistent with the reported results,[62-64] thereby once again confirming the validity of HM-PFM in ferroelectric domain characterization.

The measurement of hysteresis loop by using HM-PFM switching spectroscopy is further performed here. The switching spectroscopy in HM-PFM is similar with that from conventional PFM.[11, 44, 65] A continuous or pulse triangular wave-like DC bias sequence is superimposed on the AC drive and then applied to the sample to induce the polarization switching, while at the same time, measuring the DFP signal as a function of DC bias. Two ferroelectric materials are measured here, the first one is 300 nm Pb(Zr,Ti)O$_3$ (PZT) film, and the other is PZN-9%PT. A continuous triangular DC probing wave with a step duration of 1 ms is employed to measure the hysteresis loops, which is similar with the macroscale polarization-electric field hysteresis loop measurement.[66] Fig. 2g,h display the attained hysteresis loops of PZT and PZN-9%PT, respectively. Both the HM-PFM amplitude loops (blue curves) of PZT (Fig. 2g) and PZN-9%PT (Fig. 2h) show the expected butterfly shape which is the common characteristic of ferroelectricity.[8, 10, 66] Meanwhile, the DFP (red curves), calculated by amplitude × cos(phase), on both PZT (Fig. 2g) and PZN-9%PT (Fig. 2h) manifest the typical ferroelectric hysteresis with applied DC bias.[8, 10, 66] Therefore, the results shown in Fig. 2g,h well demonstrate that HM-PFM switching spectroscopy can provide standard hysteresis loop measurement for the study of local polarization dynamics.

**High-frequency operation.** As the HM-PFM uses the heterodyne method to detect the piezoelectric strain, the excitation frequency for sample now is no longer limited by the bandwidth of conventional optical lever system. According to the advantages provided by high-frequency excitation, one can expect a more unambiguous measurement for piezoelectric or ferroelectric information if higher frequencies are used. Herein, PPLN and PZN-9%PT are still used to explore the high-frequency detection capability of the HM-PFM. Fig. 3 displays the amplitude and phase images of the DFP signal with drive frequencies ranging from ~30 MHz to ~110 MHz. Without loss of generality, each pair of amplitude and phase images are obtained by using different tips and scanning on different areas (including both clean and contaminated areas). From the results of PPLN shown in Fig. 3a-f, clear domain wall, uniform amplitude distribution and a near 180° phase difference between the domains with opposite polarization can be observed under the excitation frequencies of $f_s$ = 28.1882 MHz (Fig. 3a,b), 40.4823 MHz (Fig. 3c,d) and 62.61 MHz (Fig. 3e,f). Similar results can also be observed on PZN-9%PT sample, which are shown in Fig. 3i,j ($f_s$ = 28.13966 MHz) and Fig. 3k,l ($f_s$ = 42.975 MHz). Surprisingly, it is found that an effective ferroelectric domain characterization for PPLN can still be achieved even when the drive frequency is increased to $f_s$ = 109.137 MHz (Fig. 3g,h). Although the amplitude and phase images shown in Fig 3g,h display a deviation from the "ideal" PFM measurements,[60] the periodical domain structure are still presented unambiguously. The deviation indicates that a background signal with the same frequency of $f_{\text{diff}}$ gets involved into the DFP signal, thus causing an obvious amplitude contrast and a non-180°phase difference between the upward and downward domains.[67] This background signal actually comes from the radio-frequency radiation effect due to the ~110 MHz high frequency and imperfect electromagnetic shielding of the holder transducer. If the holder transducer can be well electromagnetically shielded, this background signal will be minimized (see Supplementary Note 6). However, even under the influence of radio-frequency radiation, the domain structure of PPLN can still be clearly revealed by our model HM-PFM system with ordinary AFM probes at the excitation frequency up to ~110 MHz. It is still possible to further improve the operation frequency by optimizing the model HM-PFM system, such as enhancing electromagnetic shielding, using special AFM probes[68] and increasing the center frequency of the holder transducer, etc. To our best knowledge, this ~110 MHz is about 100 to 1000 times higher than the frequency used in conventional PFM and 10 times higher than the highest excitation frequency (~10 MHz[25]) used in high-frequency PFM previously.

**Electrostatic force contribution.** One of the most important considerations concerning using high frequency in PFM is to minimize the electrostatic force contribution. Herein, the electrostatic force contribution in the piezoresponse signal of HM-PFM is investigated and compared with that from conventional PFM. PPLN is still selected as the test sample, as it is reported that its PFM measurements can be affected by electrostatic force significantly.[17-19] When the sample drive voltage $V_s\sin(2\pi f_s t)$ with a DC bias $V_{dc}$ is applied between the tip and the sample, the resultant electrostatic force $F_{EF}$ is $C'[V_{dc} - V_{cpd} + V_s\sin(2\pi f_s t)]^2 / 2$, where $C'$ is the capacitance gradient of tip-sample capacitor and $V_{cpd}$ is the contact potential difference between tip and sample surface. In conventional PFM, the piezoelectric strain signal is demodulated from the first-harmonic response, thus the first-harmonic component of the electrostatic force, $F_{EF-\omega} = C'(V_{dc} - V_{cpd})V_s\sin(2\pi f_s t)$, will naturally get involved into the demodulation result. Obviously, $F_{EF-\omega}$ varies with the applied DC bias, so that the contribution of electrostatic force in PFM can be revealed by changing the DC bias. Fig. 4a,b show the single-frequency PFM amplitude and phase images of the PPLN under DC biases of 0 V and ±15 V. It is evident that when DC bias is 0 V, the expected standard amplitude and phase contrasts can be obtained. However, when DC bias is changed to ±15 V, striking amplitude contrast appeared and only slight phase difference can be observed between the domains with opposite polarization. Since ±15 V bias will not cause polarization switching due to the high coercive field of the PPLN,[69] this phenomenon is ascribed to the significant contribution from electrostatic force.[17-19] To make the comparison, exactly the same experiment is performed *in situ* by using HM-PFM and the results are displayed in Fig. 4c-f. In comparison to the remarkable change of amplitude and phase contrasts in the single-frequency PFM measurement, the results obtained by HM-PFM under two different operating frequencies ($f_s$ = 8.422 MHz in Fig. 4c,d and $f_s$ = 7.45915MHz in Fig. 4e,f) show almost no change when ±15 V DC biases are applied. To further examine the dependence of amplitude and phase with varying DC bias, the same DC spectroscopy experiments are performed at the same position by single-frequency PFM and HM-PFM. The amplitude and phase as a function of DC voltage measured by scanning the DC bias within ±10 V in single-frequency PFM are shown in Fig. 4g and 4h, respectively. Obviously, the V-shaped amplitude curve and near 180° phase change indicate that the first-harmonic electrostatic force is affecting the piezoresponse signal of PPLN significantly. By contrast, completely different variation trends of the amplitude and phase can be observed in the measurement of HM-PFM (Fig. 4i,j). It is evident that, even using various excitation frequencies, almost all of the amplitude and phase signals keep a constant magnitude with the changing DC bias. The same experiments have also been conducted on X-cut quartz single crystal additionally, and the results also show a similar constant trend (see Supplementary Fig. 6), which strongly indicates that the contribution from the electrostatic force in HM-PFM measurement has been significantly minimized.

The mechanism of minimizing electrostatic force contribution in HM-PFM can be understood from the piezoresponse signal generation process shown in Fig. 5. The tip is excited to have a vibration of $A_t\sin(2\pi f_t t + \phi_t)$ by the mechanical wave $A_{wave}\sin(2\pi f_t t + \phi_{wave})$ generated from the holder transducer. The sample vibration $A_s\sin(2\pi f_s t + \phi_s)$ is stimulated by the sample drive via inverse piezoelectric effect. Considering the influence of first-harmonic electrostatic force $C'(V_{dc} - V_{cpd})V_s\sin(2\pi f_s t)$, which causes a second tip vibration $A_{EF}\sin(2\pi f_s t + \phi_{EF})$ through the cantilever transfer function $H_{EF}(\omega)$. With these three vibration items, the time-dependent tip-sample interaction force now becomes

$$F_{ts}(z,t) = F_{ts}\left[z_0 + A_t\sin(2\pi f_t t + \phi_t) + A_{EF}\sin(2\pi f_s t + \phi_{EF}) - A_s\sin(2\pi f_s t + \phi_s)\right] \quad (7)$$

Applying the same mathematical process of Equations (4) to (6), two difference-frequency forces controlled by sample vibration ($F_{ts}(z)_{diff-s}$) and electrostatic force ($F_{ts}(z)_{diff-EF}$) will be generated, which are respectively given by:

$$F_{ts}(z)_{diff-s} = -\frac{1}{2}F_{ts}''(z_0)A_s A_t \cos\left(2\pi f_{diff} t + \phi_s - \phi_t\right) \quad (8)$$

$$F_{ts}(z)_{diff-EF} = \frac{1}{2}F_{ts}''(z_0)A_{EF} A_t \cos\left(2\pi f_{diff} t + \phi_{EF} - \phi_t\right) \quad (9)$$

$F_{ts}(z)_{diff-s}$ and $F_{ts}(z)_{diff-EF}$ then drive the cantilever to vibrate at $A_{diff-s}\sin(2\pi f_{diff} t + \phi_{diff-s})$ and $A_{diff-EF}\sin(2\pi f_{diff} t + \phi_{diff-EF})$ respectively via cantilever transfer function $H_{ts}(\omega)$. Finally, these two difference-frequency vibrations $A_{diff-s}\sin(2\pi f_{diff} t + \phi_{diff-s})$ and $A_{diff-EF}\sin(2\pi f_{diff} t + \phi_{diff-EF})$ vectorially synthesize to the final HM-PFM DFP signal $A_{diff}\sin(2\pi f_{diff} t + \phi_{diff})$. Note that the cantilever transfer functions $H_{wave}(\omega)$, $H_{EF}(\omega)$ and $H_{ts}(\omega)$ are different due to the difference of excitation schemes, which are mechanical wave, electrostatic force and local tip-sample interaction force excitations, respectively. As $A_{diff}\sin(2\pi f_{diff} t + \phi_{diff})$ is the vector sum of $A_{diff-s}\sin(2\pi f_{diff} t + \phi_{diff-s})$ and $A_{diff-EF}\sin(2\pi f_{diff} t + \phi_{diff-EF})$, $A_{diff}\sin(2\pi f_{diff} t + \phi_{diff})$ can be dominated by $A_{diff-s}\sin(2\pi f_{diff} t + \phi_{diff-s})$ (i.e., the electrostatic force contribution is negligible) if $A_{diff-s}$ is far larger than $A_{diff-EF}$. Since $A_{diff-s}\sin(2\pi f_{diff} t + \phi_{diff-s})$ and $A_{diff-EF}\sin(2\pi f_{diff} t + \phi_{diff-EF})$ are stimulated by $F_{ts}(z)_{diff-s}$ and $F_{ts}(z)_{diff-EF}$ via an identical cantilever transfer function, $A_{diff-EF}$ and $A_{diff-s}$ are actually governed by the magnitudes of $F_{ts}(z)_{diff-s}$ and $F_{ts}(z)_{diff-EF}$. Further comparing the Equations (8) and (9), it is obvious to see that the ratio of these two forces is fundamentally determined by sample voltage-induced piezoelectric vibration $A_s$ and the electrostatic force-induced tip vibration $A_{EF}$. As a consequence, by controlling the relative magnitudes of $A_s$ and $A_{EF}$, it is able to make $A_{diff-s}$ far larger than $A_{diff-EF}$ thus realizing the minimization of electrostatic force contribution. Fortunately, for a given sample drive and electrostatic force, $A_{EF}$ is governed by the cantilever transfer function $H_{EF}(\omega)$ while $A_s$ is separately determined by the piezoelectric coefficient $d_{33}$. Under high-frequency excitation, due to the significant inertial stiffening and internal damping effect, the transfer function $H_{EF}(\omega)$ will dramatically attenuate $A_{EF}$ both for on- and off-resonance states,[18, 20, 25] which has been verified by theoretical calculation and experiment (see details in Supplementary Note 4). On the contrary, the piezoelectric coefficient $d_{33}$ is not expected to decrease within MHz frequency band,[13, 60] indicating that $A_s$ can almost keep constant under MHz high-frequency excitation. This huge difference in the frequency dependences of $H_{EF}(\omega)$ and piezoelectric coefficient now allows to significantly reduce $A_{EF}$ while keeping $A_s$ unchanged (i.e., to make $A_{\text{diff-s}}$ far larger than $A_{\text{diff-EF}}$) by just increasing the excitation frequency (Supplementary Fig. 7). Typically, in HM-PFM measurement, the drive frequency of tip $f_t$ is near the eigenfrequency of a high eigenmode, while the sample drive frequency $f_s$ is set as $f_t + f_{\text{diff}}$, which is usually located within off-resonance region thus $A_{EF}$ will not get the resonance amplification (i.e., $A_{EF}$ is attenuated both by high frequency and off-resonance). Therefore, the HM-PFM amplitude and phase in Fig. 4i,j and Supplementary Fig. 6 can keep constant when electrostatic force is largely changed, this is because, under the high-frequency excitations, $A_s$ is dominantly larger than $A_{EF}$ and the contribution from electrostatic force is negligible.

For comparison, the piezoresponse signal generation mechanism in conventional PFM had also been analyzed in Supplementary Note 4. Although increasing the excitation frequency to minimize the electrostatic force contribution can also be employed in conventional PFM,[18, 25, 27] compromise must be made to avoid significant decrease of the sample vibration signal since the transfer functions $H_{EF}(\omega)$ and $H_{ts}(\omega)$ are correlated (Supplementary Fig. 5). Meanwhile, the electrostatic force consists of two parts, the distributed part along the whole cantilever and the local part around the tip apex. Increasing drive frequency in conventional PFM is mainly able to reduce the contribution from the distributed electrostatic force,[25, 27] thus this method will be effective when the distributed electrostatic force is significant. Whereas if the local part dominates, increasing drive frequency is hard to effectively minimize the electrostatic force contribution due to the correlation of $H_{EF}(\omega)$ and $H_{ts}(\omega)$ (see details in Supplementary Note 3). However, this situation is different in HM-PFM, where both distributed and local electrostatic force contributions can be largely minimized at high frequency due to the heterodyne detection scheme (Supplementary Fig. 7). In short, introducing heterodyne detection to PFM has changed the conventional piezoresponse signal generation mechanism, and breaks the direct coupling between piezoelectric and electrostatic force signals, which greatly supports to minimize the electrostatic force contribution by using high-frequency excitation.

**Difference-frequency piezoresponse frequency spectrum.** As the electrostatic force issue has been well addressed in HM-PFM, then according to the principle of HM-PFM, if an obvious DFP signal can be observed in HM-PFM, it will strongly indicate the existence of a true sample vibration. We therefore examine the DFP signal by scanning the drive frequency $f_t$ and $f_s$ simultaneously (keep $f_{\text{diff}}$ constant) to observe the amplitude peaks of the DFP signal. As the tip vibration will get resonance amplification when $f_t$ is near the eigenfrequency of each eigenmode, if sample has detectable DFP, multiple amplitude peaks will emerge when $f_t$ and $f_s$ are changed. Accordingly, we develop this approach as a special HM-PFM measurement, which is called difference-frequency piezoresponse frequency spectrum (DFPFS).[44] To test the validity of DFPFS, three types of materials are selected here, including dielectric materials ($SiO_2$ and glass), Lithium battery materials ($LiCoO_2$ and $LiMn_2O_4$) and ferroelectric materials (PPLN and PZN-9%PT), and the X-cut quartz single crystal with thickness of 0.1 mm is also tested here as a reference. Before measuring DFPFS, the tip is electrically tuned on each sample to measure the first-order contact resonance frequency $f_{CR0}$ by using conventional PFM set-up. When performing DFPFS measurement, the same tip is used for all samples, the holder drive amplitudes and sample drive amplitudes are set to be uniform respectively for all samples. Fig. 6 shows the DFPFS measured on the above 7 samples with tip frequency from 2 to 14 MHz ($f_{diff}$ is set as the $f_{CR0}$ on each sample and all the contact resonance curves are shown in Supplementary Fig. 9). Firstly, for the quartz reference sample (with $d_{33}$ = 2.3 pm/V),[13] its DFPFS shows several observable resonance-like peaks, indicating that the DFP is detectable on materials with such weak piezoelectricity. Conspicuously, the DFPFS of the dielectric and Lithium battery samples manifest a dramatic difference with that of the ferroelectric materials. For the ferroelectric samples tested here, many resonance-like peaks have emerged in the entire of the spectrums which corresponds to a strong DFP. However, there are almost no noticeable (or very weak) peaks shown in the spectrums of dielectric and Lithium battery samples, indicating that there is no DFP or the DFP is too weak to be detected. According to the signal generation mechanism of the HM-PFM, very weak DFP signal implies that the sample vibration is hard to be excited by the applied high-frequency voltages. For the dielectric glass and $SiO_2$, it is easy to understand that there is no apparent first-order electromechanical coupling in these two materials thus no sample strain and DFP will be excited. Note that although obvious contact resonance can be observed on $SiO_2$ and glass by using conventional PFM (Supplementary Fig. 9), this contact resonance actually is induced by electrostatic force instead of electromechanical coupling, thus the DFPFS results obtained here clearly indicate that pure electrostatic force can produce ideal piezoresponse in conventional PFM but cannot induce noticeable DFP in HM-PFM. Note that for the Lithium-ion battery samples, the motion of the $Li^+$ in these materials will cause the electrochemical Vegard strain thus it is expected to generate a detectable DFP if referring to previous studies by Electrochemical Strain Microscopy (ESM).[47,50] However, concerns has been raised regarding the veracity of the ESM where the formation of ionic (e.g., $Li^+$) concentration gradients is expected to be too slow to contribute to the ESM signal at typical frequencies operated in ESM (~300 kHz).[13] In addition, a newly developed metrological PFM technique, the laser doppler vibrometer PFM,[60] which allows to quantify the real sample vibration has revealed a pretty weak electrochemical strain signal on ceria where the ESM signal is strong.[13] Therefore, the true electrochemical strain may be quite small *per se* under general ~300 kHz excitation, now this small strain will be further attenuated due to the 10 to 100 times higher frequencies used in HM-PFM, hence it is reasonable that the Lithium-ion battery materials tested here show almost no (or very weak) DFP signal in the DFPFS measurement. These results indicate that the high-frequency design in HM-PFM indeed goes into effect for minimizing both the electrostatic force and electrochemical strain contributions.

Meanwhile, the DFPFS does manifest a distinct characteristic for true sample vibration, thus it can potentially be regarded as a powerful evidence for identifying the true piezoelectricity in unknown materials. We are still conducting further study on DFPFS by using more functional materials and the results will be revealed and discussed elsewhere.

Discussion

To summarize, in this study, we have introduced a new advanced PFM technique, *Heterodyne Megasonic Piezoresponse Force Microscopy or HM-PFM*, which uses $10^6$ to $10^8$ Hz high-frequency excitation and heterodyne method to measure the piezoelectric strain at nanoscale. It has been confirmed that the HM-PFM can provide standard measurements for ferroelectric domain and hysteresis loop, and an effective domain characterization with excitation frequency up to ~110 MHz has been achieved in our model system. With using high-frequency excitation and heterodyne scheme, the contribution of electrostatic force has been significantly minimized in HM-PFM. Meanwhile, the electrochemical Vegard strain is also largely attenuated by using high frequency, thus the electrochemical artifacts can be effectively reduced when using HM-PFM to study the piezoelectric or ferroelectric properties, especially for those unknown materials. Finally, the special measurement offered by HM-PFM, difference-frequency piezoresponse frequency spectrum or DFPFS, has been demonstrated and a distinct DFPFS characteristic is observed on materials with piezoelectricity. In brief, HM-PFM has simultaneously minimized the influences from multiple signal sources, thus making the target piezoelectric signal considerably purified. Given the challenges and concerns encountered in conventional PFM, especially the signal source issue, the unique advantages of HM-PFM make it an excellent candidate for the piezoelectric or ferroelectric studies where conventional PFM measurements are highly controversial. Moreover, HM-PFM successfully opens an access to high-frequency piezoelectric vibration, thus it can potentially be used to explore the unknown electromechanical coupling phenomena or ferroelectric switching processes under high-frequency electric field. As an extension, HM-PFM can be easily modified to realize the detection of high-order electromechanical coupling, such as the $2^{nd}$-order coupling of electrostriction (Supplementary Note 7).[44] Conventional PFM-based method used to measure the electrostrictive strain may suffer the effects of $2^{nd}$-harmonic electrostatic force and Joule heating,[33, 37, 39, 55] while both of these two effects can be minimized in the HM-PFM-based method thus the measurement is expected to be more unambiguous (Supplementary Note 7).

Methods

HM-PFM setup. The model HM-PFM system was established on a commercial SPM system (SPA400, Seiko Instruments Inc.). All the AC excitation signals with frequency less than 30 MHz were generated by the arbitrary waveform generator (Keysight 33522B, Keysight Technologies). For the generation of AC signals with frequency higher than 30 MHz, a programable direct digital synthesizer (AD9959, Analog Devices Inc.) equipped with a radio-frequency power amplifier was used. An embedded field programmable gate array (FPGA) controller (NI cRIO-9064, National Instruments) equipped with one digital acquisition card (NI 9775, National Instruments) was utilized to realize the signal acquisition. DC bias was generated by using a digital source meter (Keithley 2450, Tektronix Inc.), and the superposition of DC bias to AC drive was finished by a bias-tee. The demodulation of the AC signals was completed by using a lock-in amplifier (MFLI, Zurich Instruments Ltd.). The home-made analog multiplier, low pass filter and instrumentation amplifier were used to generate the reference signal for the DFP signal demodulation and amplification. All the programmable instruments were controlled by the self-developed HM-PFM control program based on LabVIEW™ and LabVIEW™ FPGA.

SPM characterization. For the characterization of ferroelectric properties, two types of conductive probes were used: Pt coated probe (240AC-PP, OPUS) with a force constant of ~2 N/m and free resonance frequency of ~70 kHz; Pt-Ir coated probe (PPP-CONTSCPt, Nanosensors) with a force constant of ~0.2 N/m and free resonance frequency of ~25 kHz. All the measurements were conducted by contact AFM mode with tip grounded in ambient environment. The drive amplitudes were typically set to 6 ~ 20 $V_{pp}$ for holder transducer and 2 ~ 10 $V_{pp}$ for sample.

Sample preparation. A commercially available periodically poled lithium niobate (PPLN) sample was used (AR-PPLN test sample, Asylum Research, Oxford Instruments), which consists of a 3 mm × 3 mm transparent die with a thickness of 0.5 mm. The PZN-9%PT single crystal (supplied by Microfine Materials Technology Pte. Ltd., Singapore) with respective orientations of $[100]^L/[010]^W/[001]^T$ was cut into small pieces and the surface of the samples was polished with SiC papers and alumina powder using water-cooled polisher. After the polishing processes, the dimension of the samples is approximately 4 mm (width) × 4 mm (length) × 0.5 mm (thickness). $SiO_2$ is the 300 nm thermal oxide layer on $Si^+$ wafer, and the glass sample is the ordinary soda lime glass coverslip with 0.1 mm thickness. $LiCoO_2$ and $LiMn_2O_4$ samples were prepared from the commercially available powders, and the powders were firstly dispersed in ethanol and then coated on $SiO_2/Si^+$ substrate. The PZT 20/80 film with thickness of 300 nm was grown in $SrRuO_3$-buffered $SrTiO_3$ substrate with (001) orientation using pulsed laser deposition (KrF excimer laser, λ = 248 nm). The $SrRuO_3$ layer (~50 nm) was firstly deposited on the $SrTiO_3$ substrate at a temperature of 680°C and an oxygen pressure of 15 Pa. Then the PZT layer was grown on top of the $SrRuO_3$ layer at a temperature of 600°C and same 15 Pa oxygen pressure. After growth, the film was cooled to room temperature at 10°C/min in an oxygen atmosphere of 1 atm.

Data availability. The authors declare that all relevant data supporting the findings of this study are available in this article and the Supplementary Information file. Any source data deemed relevant is available from the corresponding author upon request.

## Acknowledgements

This research is supported by Ministry of Education (MoE) Singapore through National University of Singapore (NUS) under the Academic Research Fund (AcRF) of R-265-000-596-112. Q. Z. greatly acknowledge the financial support from MoE Singapore through NUS under AcRF of R-265-100-596-112. Q. Z. and K. Z. gratefully acknowledge the help of Prof. Chee Kong Chui and Dr. Chng Chin Boon (NUS). Q. Z. would like to thank Prof. Huarong Zeng and Dr. Kunqi Xu (Shanghai Institute of Ceramics, Chinese Academy of Sciences) for fruitful discussion about the SPA400 AFM. We would also like to thank Prof. Li Lu (NUS) for great support of the Li-ion battery materials and SPA400 SPM in this study.

## Author contributions

Q. Z. designed and constructed the model HM-PFM system, conducted all of the experimental measurements and calculations, analyzed the results and wrote the article. H. W. prepared the PZN-9%PT sample. Q. H. and Z. F. prepared the PZT film sample. K. Z. led the project and directed all experimental research. Q. Z. and K. Z. conceived the original idea. All authors commented on the manuscript and approved its submission.

## Additional information

**Competing interests:** The authors declare no competing interests

Figures

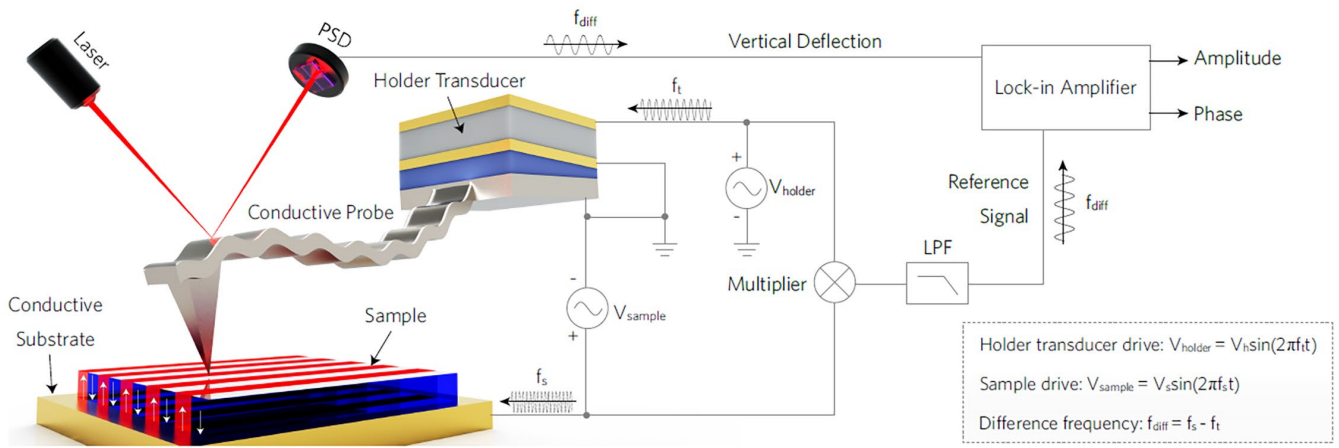

**Figure 1 | Schematic diagram of HM-PFM.** The diagram schematically shows the set-up of the model HM-PFM system (not on scale). Tip is mechanically driven by the holder transducer with a frequency $f_t$, while the sample is electrically driven by the electric field between grounded tip and conductive substrate with a frequency $f_s$. The DFP signal generated from the non-linear tip-sample interaction has a frequency of $f_{diff} = f_s - f_t$.

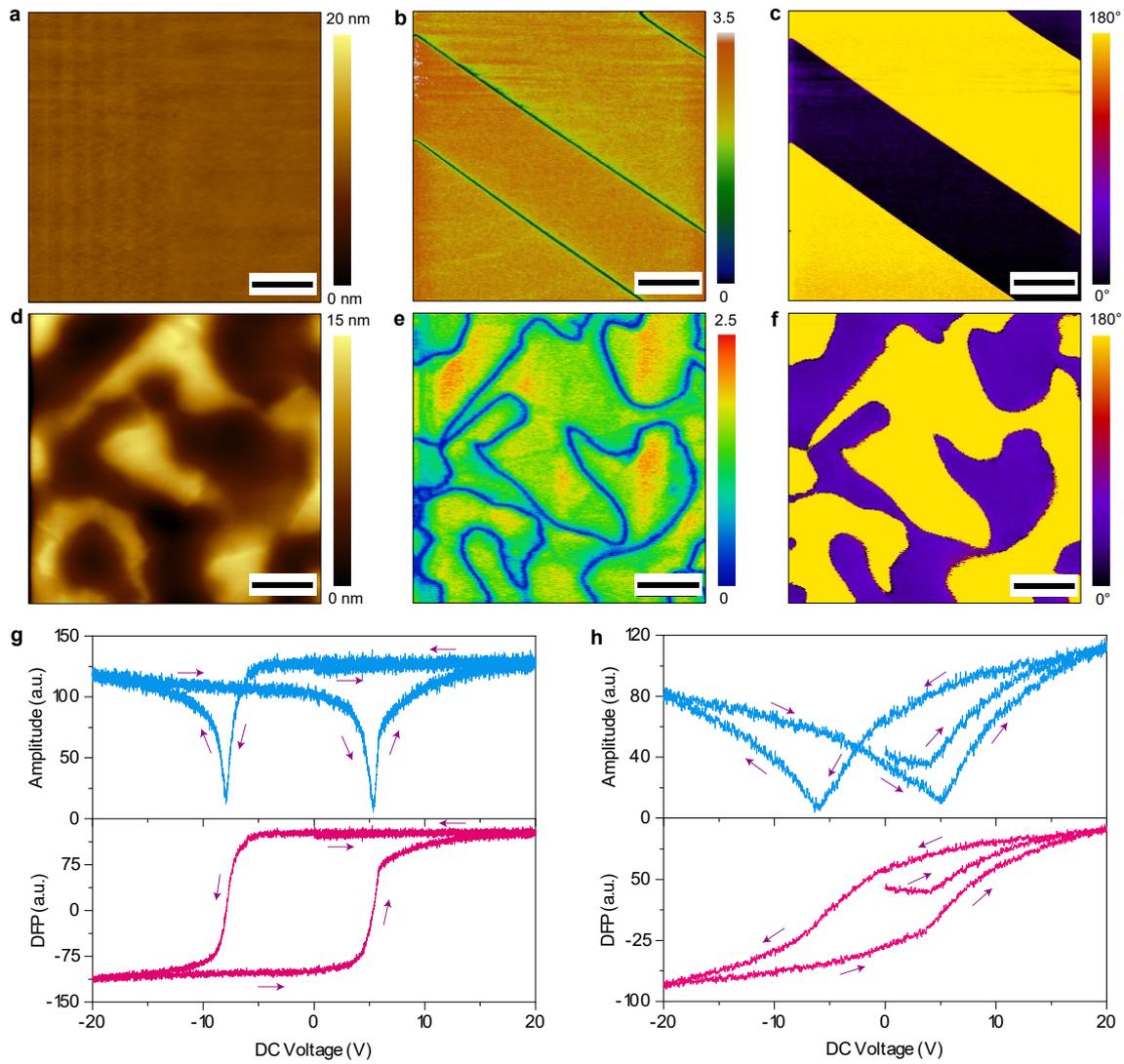

**Figure 2 | Ferroelectric domain characterization and hysteresis loop measurement.** a Topography, b HM-PFM amplitude image and c HM-PFM phase image of the PPLN sample. d Topography, e HM-PFM amplitude image and f HM-PFM phase image of the PZN-9%PT. g Hysteresis loop of PZT film and h PZN-9%PT measured by HM-PFM switching spectroscopy. Blue and red curves are amplitude and DFP (= amplitude × cos(phase)) as a function of applied DC voltage, respectively, and the arrows denote the direction of the loops. Measurement conditions: (b, c) $f_t$ = 8.356 MHz and $f_s$ = 8.63959 MHz; (e, f) $f_t$ = 8.395 MHz and $f_s$ = 8.6813 MHz; g $f_t$ = 13.516 MHz and $f_s$ = 13.82158 MHz; h $f_t$ = 15.451 MHz and $f_s$ = 15.75504 MHz. Scale bar in (a-f), 2 μm, image size = 10 × 10 μm².
22

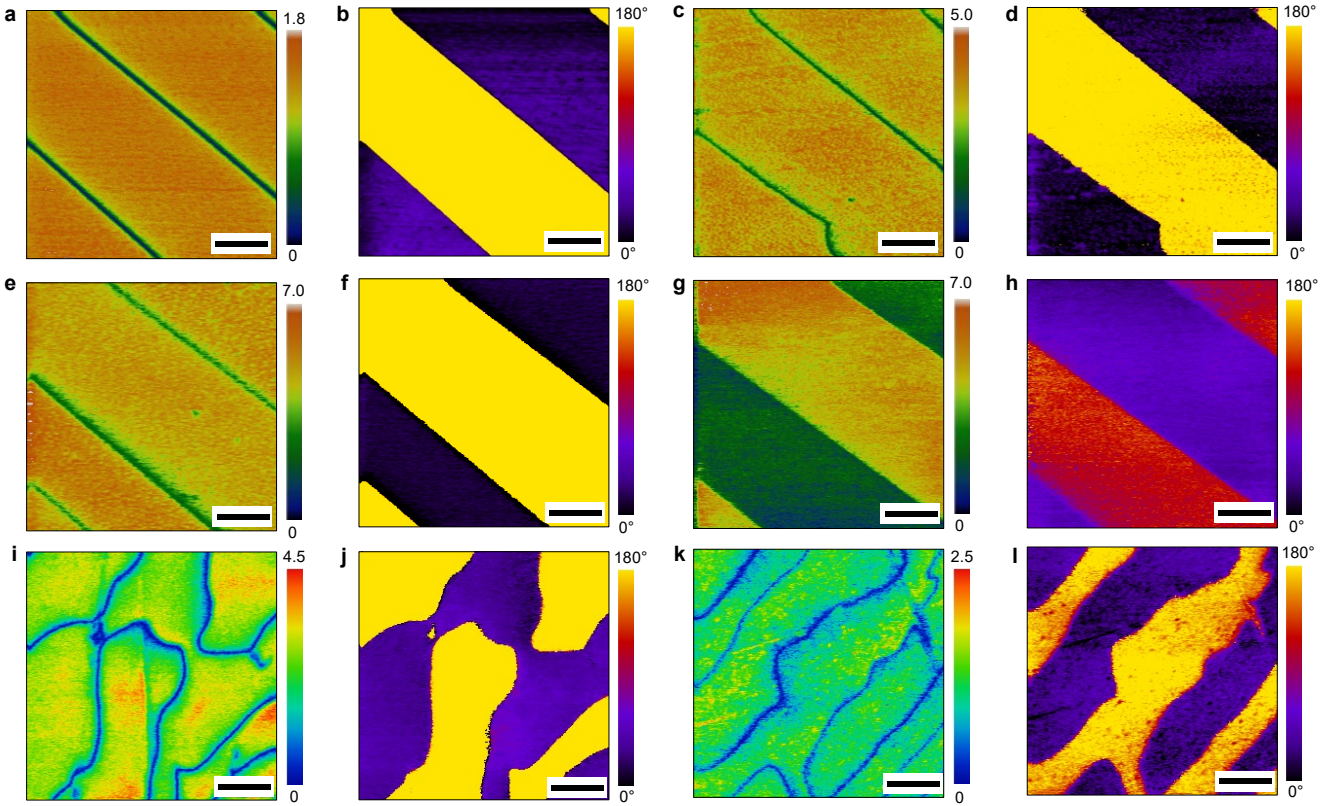

**Figure 3 | Ferroelectric domain characterization with high-frequency excitation.** (a-g) HM-PFM amplitude image (a, c, e, g) and the respective HM-PFM phase image (b, d, f, h) of PPLN sample. (i-l) HM-PFM amplitude image (i, k) and the respective HM-PFM phase image (j, l) of PZN-9%PT sample. Measurement conditions: (a, b) $f_t$ = 27.9 MHz and $f_s$ = 28.1882 MHz; (c, d) $f_t$ = 40.195 MHz and $f_s$ = 40.4823 MHz; (e, f) $f_t$ = 62.255 MHz and $f_s$ = 62.61 MHz; (g, h) $f_t$ = 108.95 MHz and $f_s$ = 109.137 MHz; (i, j) $f_t$ = 27.871 MHz and $f_s$ = 28.13966 MHz; (k, l) $f_t$ = 42.9 MHz and $f_s$ = 42.975 MHz. Scale bar in (a-g), 2 μm and image size = 10 × 10 μm$^2$; scale bar in (i-l), 1 μm and image size = 5 × 5 μm$^2$.

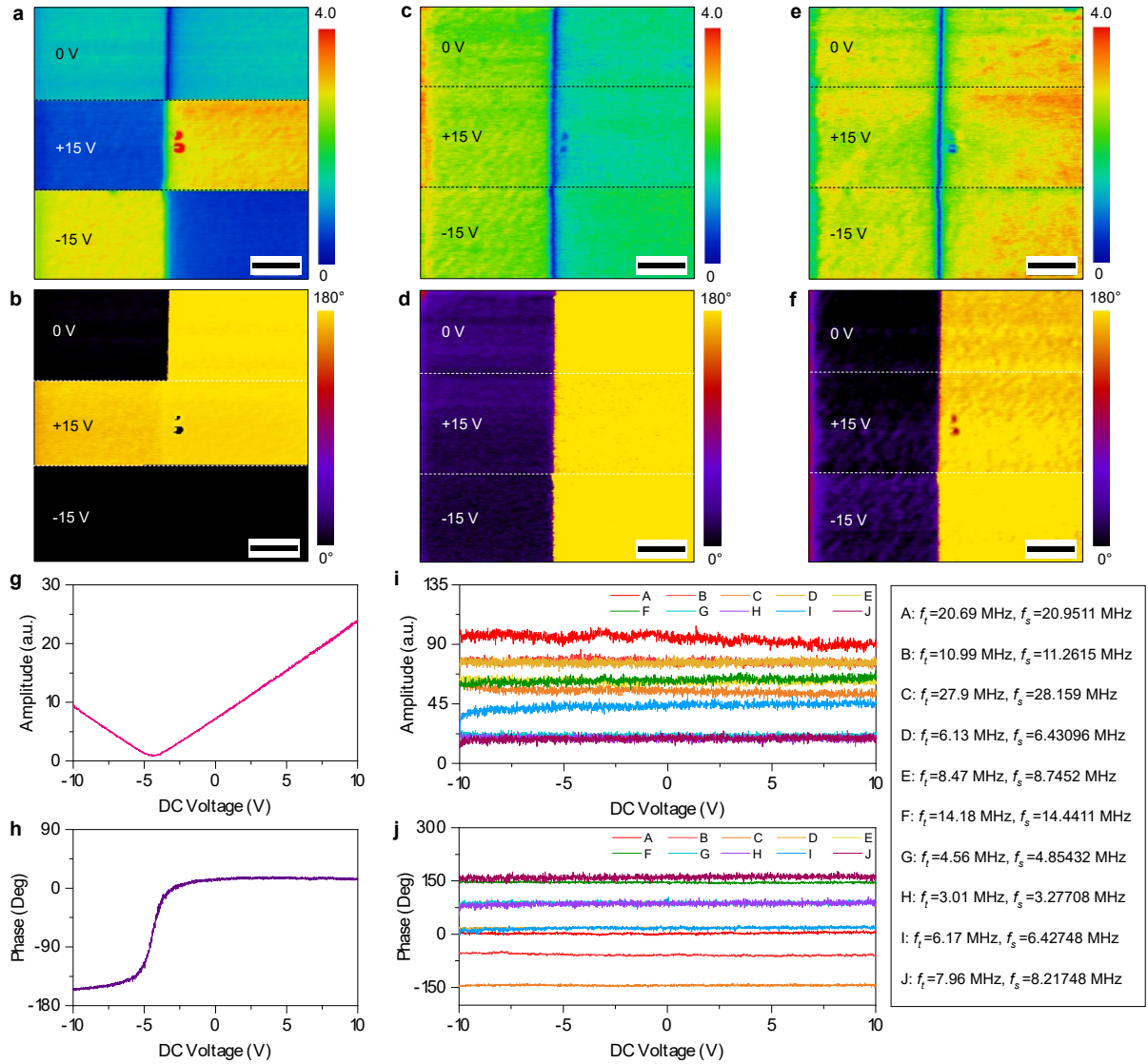

**Figure 4 | Electrostatic force contribution in the measurement of PFM and HM-PFM on PPLN.** a Single-frequency PFM amplitude image and b phase image with 0 V, +15 V and -15 V DC bias applied. (c-f) HM-PFM amplitude image (c, e) and the respective HM-PFM phase image (d, f) with 0 V, +15 V and -15 V DC bias applied. g Typical single-frequency PFM amplitude and h phase as a function of DC voltage. i HM-PFM amplitude and j phase as a function of DC voltage under various excitation frequencies (measured at the same position with g and h). Measurement conditions: (c, d) $f_t$ = 8.14 MHz and $f_s$ = 8.422 MHz; (e, f) $f_t$ = 7.176 MHz and $f_s$ = 7.45915 MHz. Scale bar in (a-f), 500 nm and image size = 3 × 3 μm$^2$.

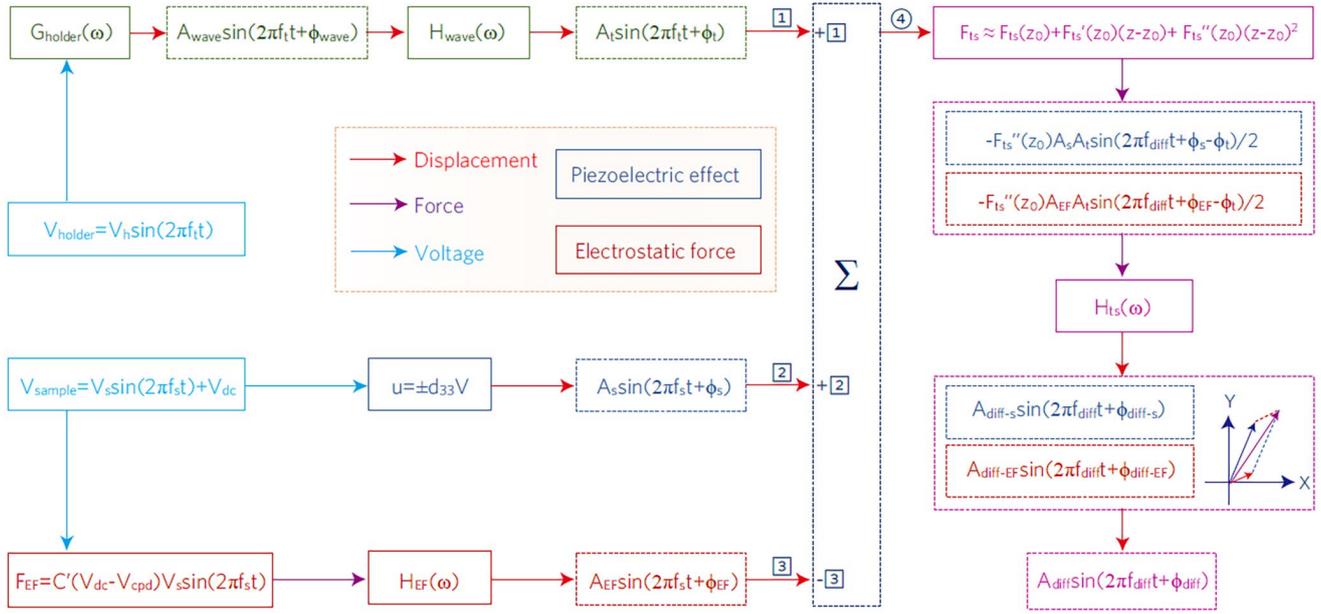

**Figure 5 | Schematic of DFP signal generation mechanism in HM-PFM with electrostatic force.** 1 Mechanically excited tip vibration $A_t\sin(2\pi f_t t + \phi_t)$, 2 sample piezoelectric vibration $A_s\sin(2\pi f_s t + \phi_s)$, 3 electrostatic force excited tip vibration $A_{EF}\sin(2\pi f_s t + \phi_{EF})$ and 4 instantaneous tip-sample separation (4 = 1 + 2 - 3). $G(\omega)$ is the transfer function of holder transducer and $u = \pm d_{33}V$ represents the linear piezoelectric effect. $H_{wave}(\omega)$, $H_{EF}(\omega)$ and $H_{ts}(\omega)$ are cantilever transfer functions under the excitations of mechanical wave, electrostatic force and local tip-sample interaction, respectively. The X-Y plane schematically depicts the vectorial synthesis of $A_{diff-s}\sin(2\pi f_{diff} t + \phi_{diff-s})$ and $A_{diff-EF}\sin(2\pi f_{diff} t + \phi_{diff-EF})$.

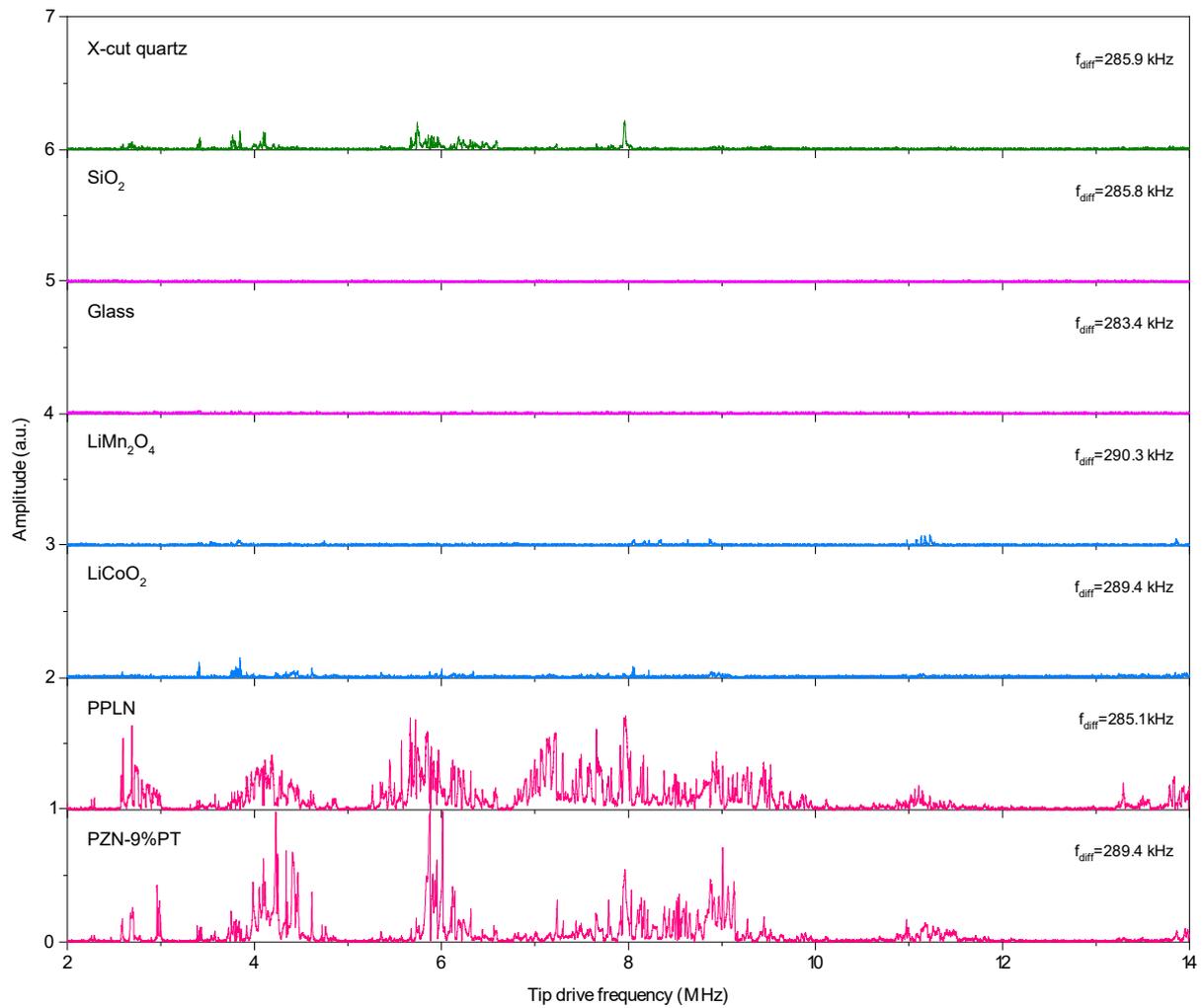

**Figure 6 | HM-PFM DFPFS measurements on different materials.** Measurement conditions: sample drive, 10 $V_{pp}$; holder transducer drive, 12 $V_{pp}$; sample drive frequency $f_s = f_t + f_{diff}$ and $f_{diff}$ is set as the first-order contact resonance frequency of the cantilever on each sample respectively. Note all the spectrums are offset for clarity.

# Supplementary Information

# A New Heterodyne Megasonic Piezoresponse Force Microscopy with High-frequency Excitation beyond 100 MHz

Qibin Zeng[1], Hongli Wang[1,3], Qicheng Huang[2], Zhen Fan[2], Kaiyang Zeng[1,*]

[1] Department of Mechanical Engineering, National University of Singapore, Singapore 117576, Singapore.

[2] Institute for Advanced Materials, South China Academy of Advanced Optoelectronics, South China Normal University, Guangzhou 510006, China.

[3] Guangdong Institute of New Materials, National Engineering Laboratory for Modern Materials Surface Engineering Technology, The Key Lab of Guangdong for Modern Surface Engineering, Guangzhou, 510651, China.

*Correspondence and requests for materials should be addressed to K. Z. (email: mpezk@nus.edu.sg)

## Supplementary Note 1. Frequency response analysis for rectangular cantilever

Supplementary Fig. 1 shows the normalized force constant $k_n$ and quality factor $Q_n$ as a function of mode number (both are normalized with respect to the first eigenmode). Compared with $Q_n$, $k_n$ shows a much faster increase with the mode number, indicating that the resonance is more and more difficult to be stimulated with increasing frequency.

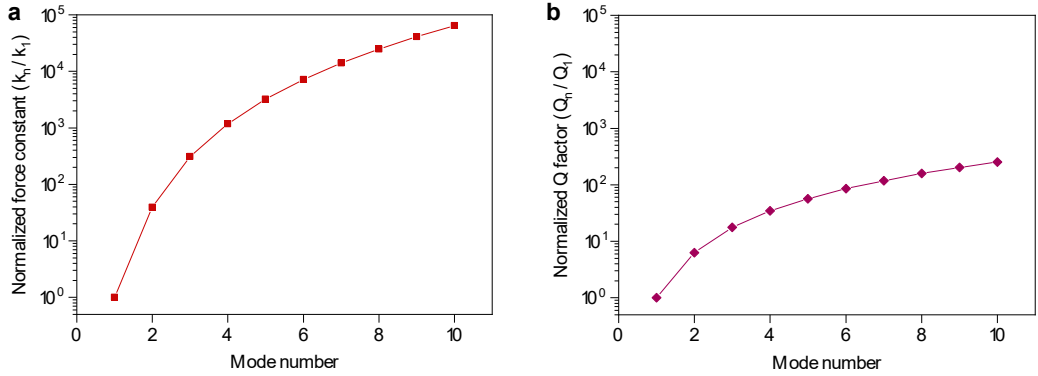

**Supplementary Fig. 1 | Dependence of force constant and Q factor with mode number.** Normalized force constant a and (b) normalized Q factor as a function of mode number, both shown logarithmically.

The amplitude response has been calculated here to show the cantilever dynamic property. According to the Euler-Bernoulli beam theory, for a rectangular cantilever, the equation of motion along its longitudinal axis is[1]

$$EI\frac{\partial^4 w(x,t)}{\partial x^4} + c\frac{\partial w(x,t)}{\partial t} + m\frac{\partial^2 w(x,t)}{\partial t^2} = 0 \quad (S1)$$

where $E$ is Young's modulus of the cantilever, $I$ is the moment of inertia, $c$ is the hydrodynamic damping coefficient, $m$ represents the mass per unit length, and $w(x, t)$ is the vertical displacement of the cantilever at the longitudinal position $x$ and time $t$. Assuming that a point force drive $f(t)$ is applied at the cantilever end ($x = L$), the boundary conditions are given by

$$w(0,t) = 0, \quad w'(0,t) = 0, \quad w''(L,t) = 0, \quad w'''(L,t) = -f(t)/EI \quad (S2)$$

Performing Laplace transform to Equation (S1) and (S2), the motion equation and boundary conditions becomes

$$EI\frac{\partial^4 W(x,s)}{\partial x^4} + (cs + ms^2)W(x,s) = 0 \quad (S3)$$

$$W(0,s) = 0, \quad W'(0,s) = 0, \quad W''(L,s) = 0, \quad W'''(L,s) = -F(s)/EI \quad (S4)$$

The solution of Equation (S3) can be expressed as[2]

$$W(x,s) = e^{\lambda x}[C_1 \sin(\lambda x) + C_2 \cos(\lambda x)] + e^{-\lambda x}[C_3 \sin(\lambda x) + C_4 \cos(\lambda x)]$$

$$\lambda = \sqrt[4]{\frac{cs + ms^2}{4EI}} \quad (S5)$$

where $C_1$, $C_2$, $C_3$ and $C_4$ are undetermined constants. The amplitude response at the cantilever end is defined as

$$A_R(s) = \frac{W(L,s)}{F(s)} \tag{S6}$$

**Supplementary Table 1 | Parameter values used in the calculation**

| Parameter | E | W | h | L | ρ | c |
|---|---|---|---|---|---|---|
| Value | 179 GPa | 40 μm | 2.6 μm | 240 μm | 2330 kg/m$^3$ | 5.6×10$^{-4}$ kg/(m.s) |

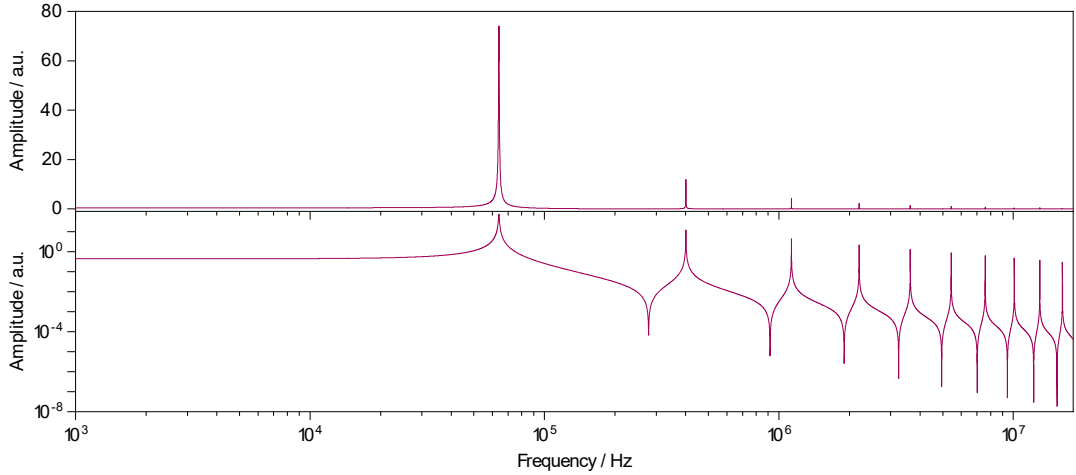

**Supplementary Fig. 2 | Amplitude response at *x=L* as a function of excitation frequency.** The amplitude is shown in linear (up) and logarithmic (down) scales respectively.

Then using the parameters from AFM tip (240AC-PP, OPUS) and a hydrodynamic damping of $5.06 \times 10^{-4}$ kg/(m·s)$^3$ (shown in Supplementary Table 1) to solve the Equation (S3) to (S6) analytically using MATLAB, the amplitude response $A_R(s)$ can be obtained and its magnitude in frequency domain is plotted in Supplementary Fig. 2.

## Supplementary Note 2. Heterodyne detection principle of HM-PFM

Considering both tip and sample surface have a vibration, $A_t\sin(\omega_t t + \phi_t)$ and $A_s\sin(\omega_s t + \phi_s)$ respectively (here use the angular frequency, $\omega_t = 2\pi f_t$, $\omega_s = 2\pi f_s$, and similarly hereinafter), the time-dependent tip-sample interaction force is given by

$$F_{ts}(z,t) = F_{ts}\left[z_0 + A_t \sin(\omega_t t + \phi_t) - A_s \sin(\omega_s t + \phi_s)\right] \tag{S7}$$

Expanding the tip-sample interaction with a Taylor series at $z=z_0$ up to second order gives

$$F_{ts}(z,t) \approx F_{ts}(z_0) + F'_{ts}(z_0)(z-z_0) + \frac{1}{2}F''_{ts}(z_0)(z-z_0)^2$$
$$= F_{ts}(z_0) + F'_{ts}(z_0)A_t \sin(\omega_t t + \phi_t) - F'_{ts}(z_0)A_s \sin(\omega_s t + \phi_s)$$
$$+ \frac{1}{2}F''_{ts}(z_0)A_t^2 \sin^2(\omega_t t + \phi_t) + \frac{1}{2}F''_{ts}(z_0)A_s^2 \sin^2(\omega_s t + \phi_s)$$
$$- F''_{ts}(z_0)A_t A_s \sin(\omega_t t + \phi_t)\sin(\omega_s t + \phi_s)$$
(S8)

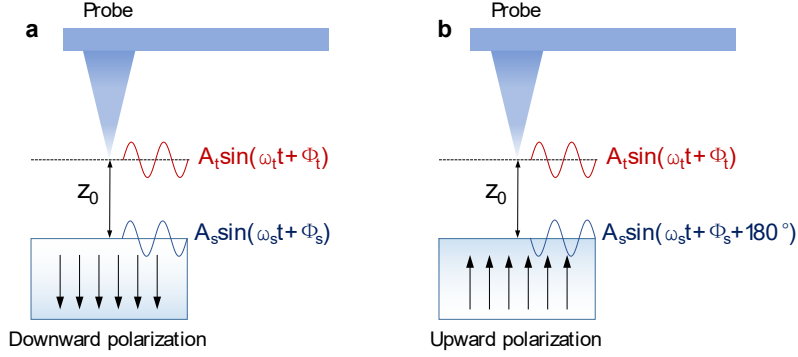

**Supplementary Fig. 3 | Schematic of the tip-sample interactions on different ferroelectric domains.** a Dynamic tip-sample interaction on downward and b upward polarization domain.

To further process the Equation (S8), the tip-sample interaction force can be expressed by the sum of the components with different frequencies

$$F_{ts}(z,t) \approx F_{ts}(z)_{static} + F_{ts}(z)_{1\omega} + F_{ts}(z)_{2\omega} + F_{ts}(z)_{sum} + F_{ts}(z)_{diff} \quad (S9)$$

where the $F_{ts}(z)_{static}$, $F_{ts}(z)_{1\omega}$, $F_{ts}(z)_{2\omega}$, $F_{ts}(z)_{sum}$ and $F_{ts}(z)_{diff}$ represent the static, first harmonic, second harmonic, sum-frequency and difference-frequency components of the tip-sample interaction force, respectively, which are given by

$$\begin{cases} F_{ts}(z)_{static} = F_{ts}(z_0) + \frac{1}{4}F''_{ts}(z_0)(A_t^2 + A_s^2) \\ F_{ts}(z)_{1\omega} = F'_{ts}(z_0)A_t \sin(\omega_t t + \phi_t) - F'_{ts}(z_0)A_s \sin(\omega_s t + \phi_s) \\ F_{ts}(z)_{2\omega} = -\frac{1}{4}F''_{ts}(z_0)A_t^2 \cos(2\omega_t t + 2\phi_t) - \frac{1}{4}F''_{ts}(z_0)A_s^2 \cos(2\omega_s t + 2\phi_s) \\ F_{ts}(z)_{sum} = \frac{1}{2}F''_{ts}(z_0)A_t A_s \cos[(\omega_t + \omega_s)t + \phi_t + \phi_s] \end{cases} \quad (S10)$$

$$F_{ts}(z)_{diff} = -\frac{1}{2}F''_{ts}(z_0)A_t A_s \cos(\omega_{diff} t + \phi_s - \phi_t) \quad (S11)$$

in which $\omega_{diff} = \omega_s - \omega_t$ is the difference frequency. For ferroelectric domains with upward and downward polarization, there is a 180° phase difference between the tip voltage-induced piezoelectric strain, thus the vibration of sample surface has a 180° phase difference between up- and downward domains,[4] which are schematically shown in Supplementary Fig. 3. From Equation (S11), when the drive signals applied to the tip and sample remain constant, the difference-frequency tip-sample interaction force on upward and downward ferroelectric domains are calculated by

$$F_{ts}(z)_{diff-down} = -\frac{1}{2}F_{ts}''(z_0)A_t A_s \cos(\omega_{diff}t + \phi_s - \phi_t) \tag{S12}$$

$$F_{ts}(z)_{diff-up} = -\frac{1}{2}F_{ts}''(z_0)A_t A_s \cos(\omega_{diff}t + \phi_s - \phi_t + 180°) \tag{S13}$$

Obviously, there is a theoretical 180° phase difference between $F_{ts}(z)_{diff-up}$ and $F_{ts}(z)_{diff-down}$, implying that the heterodyne-based PFM has the same capability with conventional vertical PFM with respect to characterizing the polarization of ferroelectric domain.

## Supplementary Note 3. Transfer functions of cantilever

To compare the difference of cantilever dynamics under electrostatic force and sample vibration excitations, the cantilever transfer functions have been calculated here. Since the AFM tip is always in contact with sample surface during PFM measurements, the vertical tip-sample coupling is modelled as a spring in parallel with a dashpot (Kelvin-Voigt model) and no lateral contact coupling is considered for simplicity.[5, 6] Supplementary Fig. 4 shows the mechanical models used in these calculations. Considering there is only a sample vibration $u(t)$ (Supplementary Fig. 4b), the cantilever is driven by the local tip-sample interaction, the equation of motion and the corresponding boundary conditions are:[6]

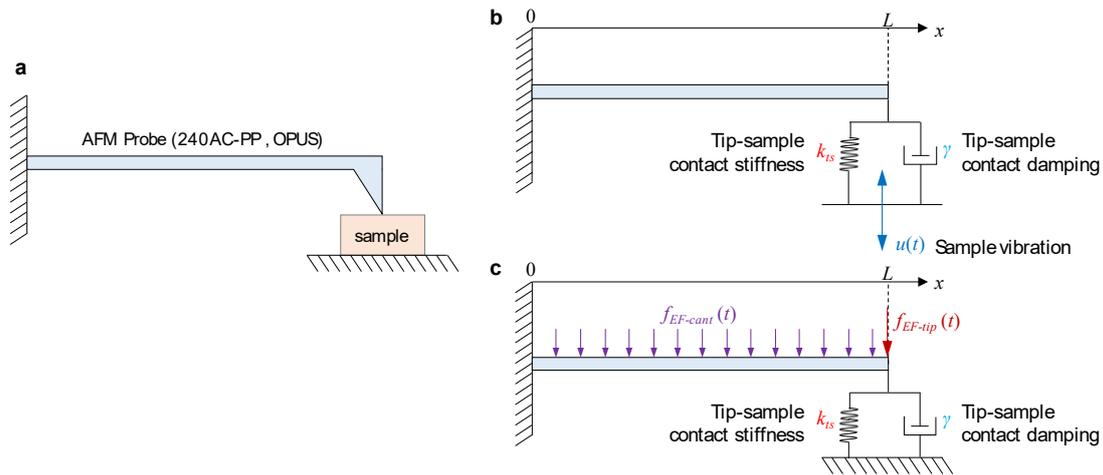

**Supplementary Fig. 4 | Mechanical models of AFM probe in contact with the sample.** a Schematic illustration of the AFM probe in contact with the sample. Mechanical models for the calculation of cantilever transfer function with the excitation of b sample vibration and c electrostatic force.

$$\begin{cases} EI\dfrac{\partial^4 w(x,t)}{\partial x^4}+c\dfrac{\partial w(x,t)}{\partial t}+m\dfrac{\partial^2 w(x,t)}{\partial t^2}=0 \\ w(0,t)=0,\ w'(0,t)=0,\ w''(L,t)=0,\ w'''(L,t)=\dfrac{1}{EI}\left[k_{ts}\big(w(L,t)-u(t)\big)+\gamma\dfrac{\partial(w(L,t)-u(t))}{\partial t}\right] \end{cases} \quad (S14)$$

in which $k_{ts}$ and $\gamma$ are tip-sample contact stiffness and contact damping constant, respectively. Performing Laplace transform to Equation (S14),

$$\begin{cases} EI\dfrac{\partial^4 W(x,s)}{\partial x^4}+(cs+ms^2)W(x,s)=0 \\ W(0,s)=0,\ W'(0,s)=0,\ W''(L,s)=0,\ W'''(L,s)=\dfrac{(k_{ts}+\gamma s)[W(L,s)-U(s)]}{EI} \end{cases} \quad (S15)$$

and defining the transfer function for sample vibration (also the local tip-sample interaction) excitation as

$$H_{ts}(s)=\dfrac{W(L,s)}{U(s)} \quad (S16)$$

The solution of Equation (S15) has exactly the same form of Equation (S5), thus by solving Equation (S15), (S5) and (S16) analytically in MATLAB, the transfer function in Laplace domain $H_{ts}(s)$ can be obtained and the frequency domain form $H_{ts}(\omega)$ can be transformed from $H_{ts}(s)$ via $s = i\omega$. Using the parameters shown in Table 1 and assuming $k_{ts} = 100k_c = 100 \times 3EI/L^3$, $\gamma = 0.1 \times (EIm)^{0.5}/L$ (dimensionless damping constant of 0.1)[7], the calculated transfer function $H_{ts}(\omega)$ is plotted in Supplementary Fig. 5.

For the electrostatic force, it is usually considered as the sum of two parts, the distributed part along the whole cantilever $f_{EF\text{-}cant}$ and the local part around the tip apex $f_{EF\text{-}tip}$[4, 5, 8] (Supplementary Fig. 4c), thus under the electrostatic force excitation, the equation of motion and the corresponding boundary conditions now become

$$\begin{cases} EI\dfrac{\partial^4 w(x,t)}{\partial x^4}+c\dfrac{\partial w(x,t)}{\partial t}+m\dfrac{\partial^2 w(x,t)}{\partial t^2}=\dfrac{f_{EF-cant}(t)}{L} \\ w(0,t)=0,\ w'(0,t)=0,\ w''(L,t)=0,\ w'''(L,t)=\dfrac{1}{EI}\left[k_{ts}w(L,t)+\gamma\dfrac{\partial w(L,t)}{\partial t}-f_{EF-tip}(t)\right] \end{cases} \quad (S17)$$

Performing Laplace transform to Equation (S17),

$$\begin{cases} EI\dfrac{\partial^4 W(x,s)}{\partial x^4}+(cs+ms^2)W(x,s)=\dfrac{F_{EF-cant}(s)}{L} \\ W(0,s)=0,\ W'(0,s)=0,\ W''(L,s)=0,\ W'''(L,s)=\dfrac{(k_{ts}+\gamma s)W(L,s)-F_{EF-tip}(s)}{EI} \end{cases} \quad (S18)$$

and a particular solution of Equation (S18) is $\dfrac{F_{EF-cant}(s)}{L(cs+ms^2)}$,[2] thus the solution of Equation (S18) can be written as

$$W(x,s)=e^{\lambda x}\left[C_1\sin(\lambda x)+C_2\cos(\lambda x)\right]+e^{-\lambda x}\left[C_3\sin(\lambda x)+C_4\cos(\lambda x)\right]+\dfrac{F_{EF-cant}(s)}{L(cs+ms^2)} \quad (S19)$$

Since the local and distributed electrostatic force are correlated, here for simplicity, a constant ratio $\alpha$ is introduced to relate $f_{EF\text{-}cant}$ and $f_{EF\text{-}tip}$

$$f_{EF-tip}(s) = \alpha f_{EF-cant}(s), \quad F_{EF-tip}(s) = \alpha F_{EF-cant}(s) \tag{S20}$$

then the transfer function for the electrostatic force excitation can be defined as

$$H_{EF}(s) = \frac{W(L,s)}{F_{EF-cant}(s)} \tag{S21}$$

Using the same parameters and calculation method mentioned above to solve Equation (S18) to (S21), the transfer function in frequency domain $H_{EF}(\omega)$ can be obtained. As the relative magnitude between $f_{EF\text{-}cant}$ and $f_{EF\text{-}tip}$ depends on specific experiment,[4, 9] three values of $\alpha$, 0.001, 0.1 and 10, are used to calculate $H_{EF}(\omega)$ and the results are all plotted in Supplementary Fig. 5.

Comparing the transfer function $H_{EF}(\omega)$ (dot line in Supplementary Fig. 5) and $H_{ts}(\omega)$ (red solid line in Supplementary Fig. 5), it is clear to see that both curves show a decay trend with increasing frequency, while $H_{EF}(\omega)$ obviously decreases faster than $H_{ts}(\omega)$, thus using high frequency in conventional PFM can minimize the contribution of electrostatic force. Furthermore, when $\alpha$ is small, i.e. the electrostatic force is dominated by the distributed part $f_{EF\text{-}cant}$, $H_{EF}(\omega)$ (green dot line) shows a large difference with $H_{ts}(\omega)$ and it decays much faster than $H_{ts}(\omega)$, indicating that high frequency can largely minimize the electrostatic force contribution. However, when $\alpha$ is large, i.e. local part $f_{tip\text{-}cant}$ dominates the electrostatic force, $H_{EF}(\omega)$ (black dot line) is very closed to $H_{ts}(\omega)$, this is because $f_{tip\text{-}cant}$ and sample vibration both belong to local tip-sample interaction excitation. Therefore, if $f_{tip\text{-}cant}$ dominates the electrostatic force, using high frequency in conventional PFM cannot effectively minimize the electrostatic force contribution as the target piezoresponse signal varies synchronously with electrostatic force signal.

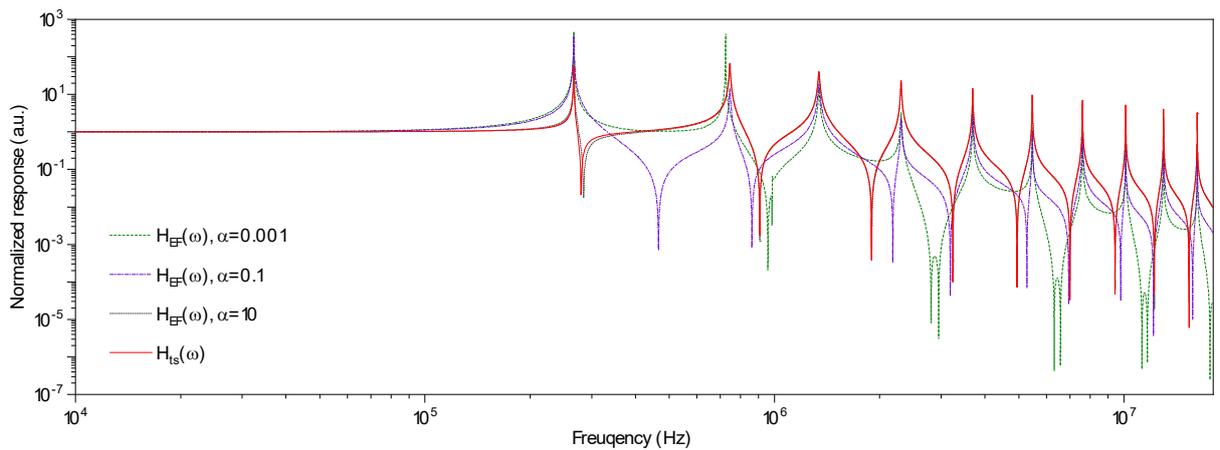

**Supplementary Fig. 5 | Calculated transfer functions of cantilever.** The magnitudes of the transfer functions are all normalized with respect to the low-frequency responses respectively.

## Supplementary Note 4. Principle of minimizing the contribution of electrostatic force

X-cut quartz single crystal with thickness of 0.1 mm has been tested here to examine the contribution of electrostatic force in HM-PFM. By scanning DC bias within ±10 V, the amplitude and phase of the difference-frequency piezoresponse (DFP) signal under various excitation frequencies are recorded and the results are all shown in Supplementary Fig. 6. Similar with the Fig. 4i,j in main text, here all of the amplitude and phase signals also keep constant with changing DC bias, implying that, even on material with weak piezoelectricity, the contribution of electrostatic force can still be neglected in HM-PFM.

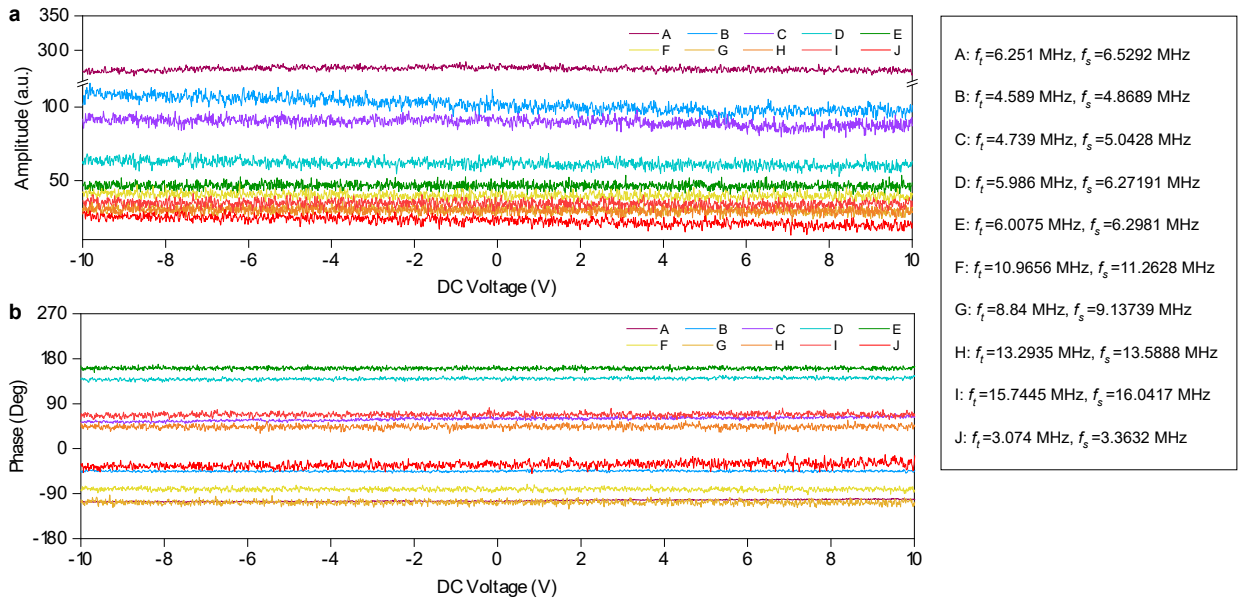

**Supplementary Fig. 6 | HM-PFM DC spectroscopy of X-cut quartz single crystal.** a Amplitude and b phase of the DFP signal as a function of DC bias under varies drive frequencies.

The key factor that enables the minimization of electrostatic force contribution in HM-PFM is the difference between frequency dependences of cantilever transfer function $H_{ts}(\omega)$ and piezoelectric strain. Here, still using the same parameters and calculation method described in Supplementary Note 3, and assuming the total electrostatic force is 10 nN,[4] the amplitude of the electrostatic force excited tip vibration $A_{EF}$ can be calculated. Supplementary Fig. 7a shows the calculated $A_{EF}$ as a function of frequency under $\alpha$ = 0.1 and 10. For a typical PFM measurement on PPLN sample, the amplitude of the sample vibration $A_s$ is ~10 pm, as the piezoelectric strain does not change apparently within MHz frequency region, the amplitude $A_s$ is schematically plotted by a horizontal line in Supplementary Fig. 7a. By comparing $A_{EF}$ and $A_s$, it is obvious that there exists a huge difference between their frequency dependences, $A_{EF}$ gradually decays with increasing excitation frequency, and in particular, $A_{EF}$ attenuates dramatically at off-resonance or anti-resonance states (such as the shadow area). For low frequency, $A_{EF}$ is larger than $A_s$, implying that the electrostatic force contribution is significant or even dominant. But at high frequency, by properly choosing the frequency (e.g. in the shadow area), $A_{EF}$ (with $\alpha = 0.1$ and 10 both) can be largely attenuated to be much smaller than $A_s$, thus realizing significant minimization of electrostatic force contribution no matter the electrostatic force is dominated by distributed or local part. Note that the calculation of $A_{EF}$ here is based on multiple assumptions, such as ignoring the internal damping and the frequency dependence of contact damping, the practical attenuation of $A_{EF}$ is much more rapid. Supplementary Fig. 7b shows the experimentally measured tip vibration amplitude (the amplitude of cantilever deflection signal) as a function of excitation frequency. This curve is measured on clean $SiO_2$ surface by conventional PFM set-up with the same AFM tip used for the calculation (i.e., 240AC-PP).

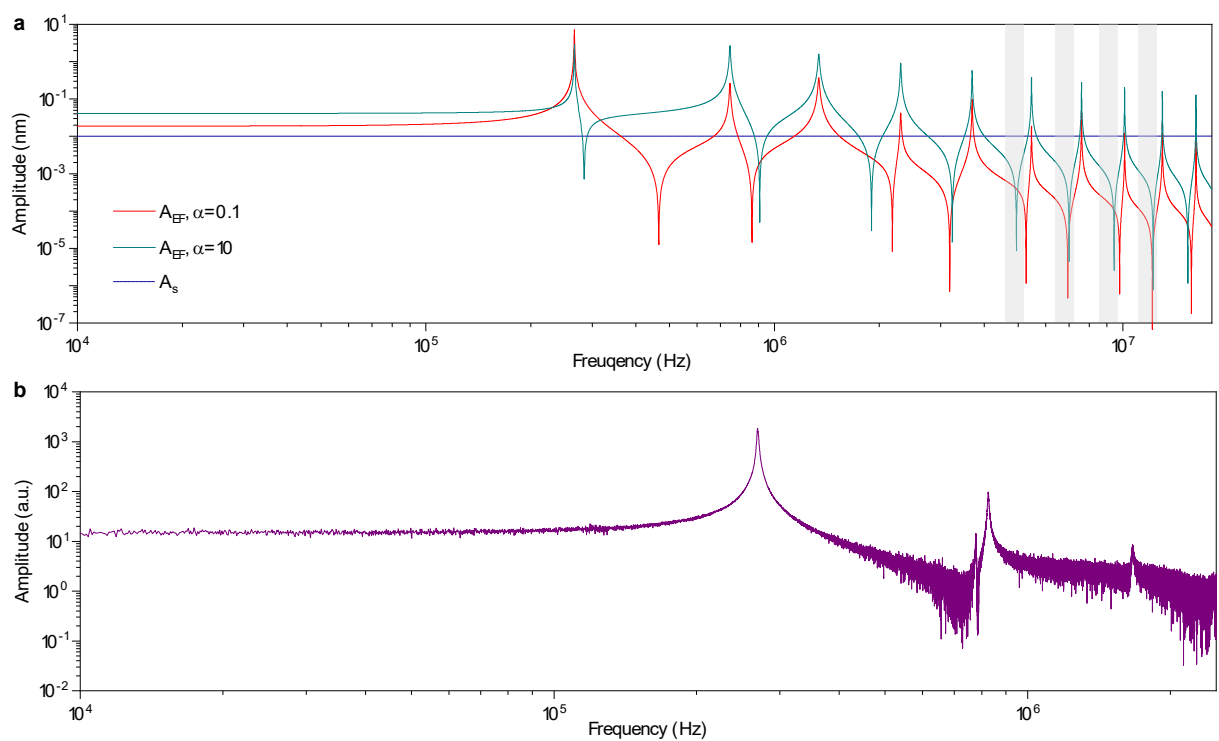

**Supplementary Fig. 7 | Electrostatic force excited amplitude as a function of excitation frequency.** a Calculated amplitude $A_{EF}$ (at $x=L$) as a function of excitation frequency, a piezoelectric vibration with amplitude $A_s$=10 pm is schematically shown by the horizontal dot line. b Amplitude of the deflection signal as a function of excitation frequency (measured by single-frequency PFM on $SiO_2$ Sample).

As the $SiO_2$ is the pure dielectric layer of Si wafer, the measured tip vibration is actually stimulated by the electrostatic force. It is obvious that the practical electrostatic force excited tip vibration decays very fast, just from the 1st to 3rd eigenmode, the resonant amplitude has already attenuated ~100 times, which highly indicates that at much higher frequency, the practical $A_{EF}$ will further decrease to be far smaller than $A_s$. Therefore, HM-PFM can achieve almost an ideal minimization of electrostatic force contribution.

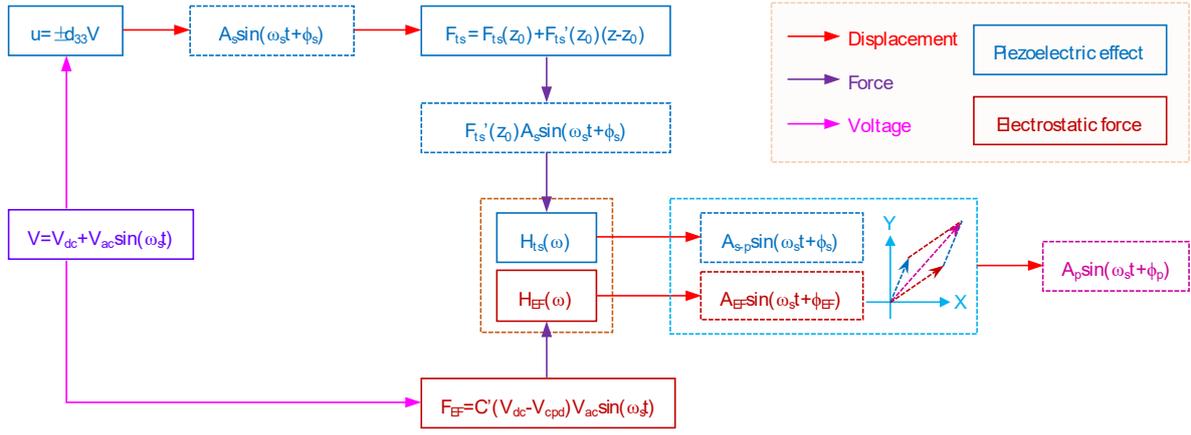

**Supplementary Fig. 8 | Schematic of piezoresponse signal generation mechanism in conventional PFM.**

Supplementary Fig. 8 shows that, in the conventional PFM, the tip voltage stimulated piezoelectric vibration will cause a local varying tip-sample interaction force $F_{ts}'(z_0)\sin(\omega_s t + \phi_s)$, this force will drive the cantilever to vibrate and generate the piezoelectric signal $A_{s-p}\sin(\omega_s t + \phi_{p-s})$ via the cantilever's transfer function $H_{ts}(\omega)$. At the same time, the first harmonic electrostatic force $C'(V_{dc} - V_{cpd})V_{ac}\sin(\omega_s t)$ directly drive the cantilever to vibrate, generating the electrostatic force signal $A_{EF}\sin(\omega_s t + \phi_{p-EF})$ via transfer function $H_{EF}(\omega)$. Finally, the piezoelectric signal $A_{s-p}\sin(\omega_s t + \phi_{p-s})$ and electrostatic force signal $A_{EF}\sin(\omega_s t + \phi_{p-EF})$ will vectorially synthesis to the final piezoresponse signal $A_p\sin(\omega_s t + \phi_p)$. Although multiple methods, such as using probes with large force constant and operating at high frequency or higher eigenmodes,[5, 10-12] are proposed to minimize the contribution of electrostatic force, these methods in principle are based on the difference between $H_{ts}(\omega)$ and $H_{EF}(\omega)$. However, as shown in Supplementary Fig. 5, $H_{ts}(\omega)$ and $H_{EF}(\omega)$ belong to the same cantilever and are correlated internally, it is almost impossible to change $H_{EF}(\omega)$ only while keeping $H_{ts}(\omega)$ unaffected especially when the electrostatic force is dominated by the local part. Therefore, compromise must be made to avoid significant damage to the target piezoelectric signal when addressing the electrostatic force issue in conventional PFM. Similarly, as the Electrochemical Strain Microscopy (ESM) has exactly the same set-up with that of the conventional PFM, the signal generation mechanism discussed above is also applied to ESM where the electrostatic force is also an important issue.[13, 14]

## Supplementary Note 5. Contact resonance curves measured by single-frequency PFM

Supplementary Fig. 9 shows the contact resonance curves of the 7 samples discussed in the Fig. 6 of the main text, in which are all measured by single-frequency PFM with sample drive amplitude of 2 ~ 10 $V_{pp}$. The resonance frequencies are all fitted from the resonance curves by using simple harmonic oscillator model.[15]

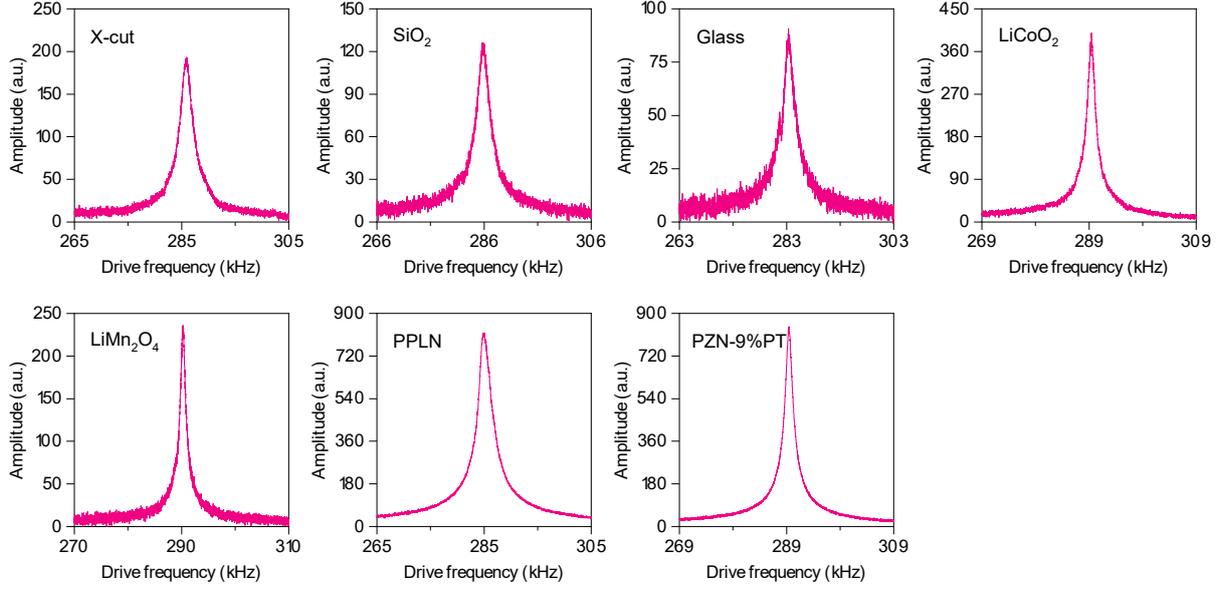

**Supplementary Fig. 9 | Contact resonance curves measured by single-frequency PFM.**

Supplementary Note 6. Influence of the radio-frequency radiation

With increasing the excitation frequency of the holder transducer, the radio-frequency radiation around the transducer increases, causing an additional electric field $E_{RF}$ between the tip and substrate. This radiated electric field $E_{RF}$ has the same frequency with the transducer drive. Assuming that $E_{RF}$ is caused by an equivalent AC voltage $V_{RF}\sin(\omega_t t)$ applied between tip and substrate, the total voltage between tip and substate is (ignoring the DC bias and phase here)

$$V_{total} = V_s \sin(\omega_s t) + V_{RF} \sin(\omega_t t) \quad (S22)$$

Then the total electrostatic force under the radiation is given by:

$$F_{EF}^* = \frac{1}{2}C'(V_{total} - V_{cpd})^2 \quad (S23)$$

Substituting Equation (S22) into (S23) and rearranging:

$$\begin{aligned} F_{EF}^* = &\frac{1}{2}C'\left(V_{cpd}^2 + \frac{1}{2}V_s^2 + \frac{1}{2}V_{RF}^2\right) \\ &-\frac{1}{2}C'V_{cpd}\left[V_s \sin(\omega_s t) + V_{RF} \sin(\omega_t t)\right] \\ &-\frac{1}{4}C'\left[V_s^2 \cos(2\omega_s t) + V_{RF}^2 \cos(2\omega_s t)\right] \\ &-\frac{1}{4}C'V_s V_{RF} \cos\left[(\omega_s + \omega_t)t\right] \\ &+\frac{1}{4}C'V_s V_{RF} \cos(\omega_{diff} t) \end{aligned} \quad (S24)$$

From Equation (S24), it is evident that there is a difference-frequency component in $F_{EF}^*$, $C'V_s V_{RF}\cos(\omega_{diff}t)/4$, which has the same frequency with DFP signal thus it can drive the cantilever and affect the HM-PFM results. However, if the radiation is effectively shielded, $V_{RF}$ will be closed to zero thus the contribution from the difference-frequency electrostatic force can be neglected.

## Supplementary Note 7. Measurement of electrostriction

The HM-PFM developed here can be easily modified to measure the high-order electromechanical coupling, such as the important 2$^{nd}$-order coupling of electrostriction. Conventional PFM-based method has been extensively used to study the electrostriction,[16-19] while this method is typically influenced by the 2$^{nd}$-harmonic electrostatic force and Joule heating. Obviously, the 2$^{nd}$-harmonic electrostatic force contribution can be largely minimized in HM-PFM-based method by similar means of the 1$^{st}$-harmonic electrostatic force. Then due to the periodical temperature variation decays with increasing frequency,[20] the Joule heating-induced thermal strain can also be diminished in HM-PFM-based measurement. Supplementary Fig. 10 shows the modified HM-PFM set-up for the measurement of electrostriction,[21] in which the main modification is the reference signal generation circuit. However, if the reference signal is provided internally by synchronizing the clocks of signal source and lock-in amplifier, no hardware change is needed. As the electrostriction is a quadratic effect, the target electrostrictive vibration locates at the 2$^{nd}$-harmonic of the sample drive. Therefore, after heterodyne process, the electrostrictive vibration information is included in the cantilever deflection signal with frequency of $2f_s - f_t$. In a similar fashion, the 3$^{rd}$-order electromechanical coupling information should reside in the signal with frequency of $3f_s - f_t$ and the $n^{th}$-order is in $nf_s - f_t$. If the clocks of signal source and lock-in amplifier can be synchronized, then detecting these high-order electromechanical couplings by HM-PFM-based method will be quite straightforward.

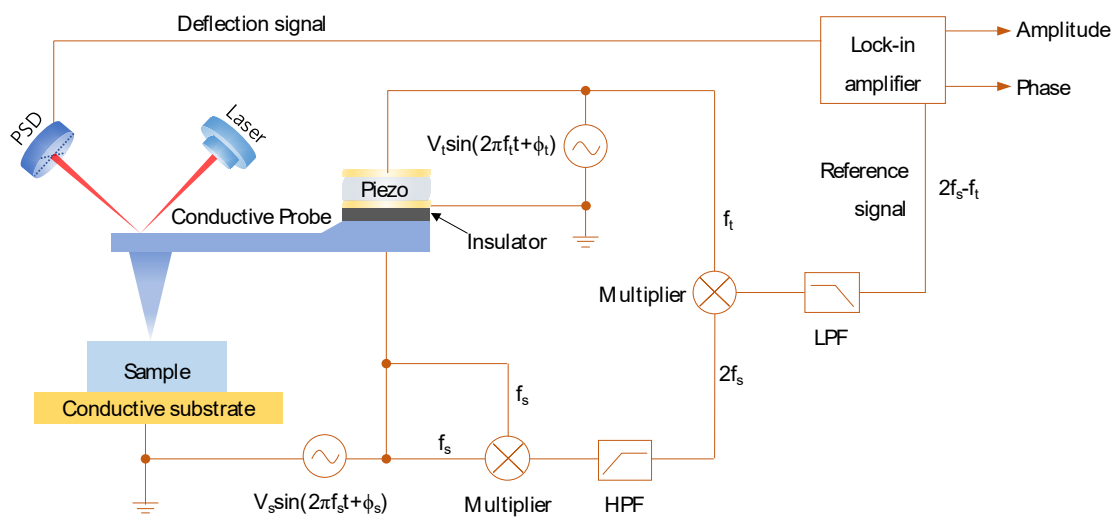

Supplementary Fig. 10 | Schematic of the modified HM-PFM set-up for the measurement of electrostriction.